RESEARCH ARTICLE

# Logarithmic-scale variational quantum eigensolver for off-lattice protein structure prediction in continuous torsional angle space

Fabio Cumbo [1,†], Bryan Raubenolt [1,†], Varun Puram [1,†], Natalie Katzenmeyer [1], Jayadev Joshi [1], Daniel Blankenberg [1,2,*]

[1] Computational Life Sciences, Cleveland Clinic Research, Cleveland Clinic, Cleveland, OH 44195, USA
[2] Department of Molecular Medicine, Cleveland Clinic Lerner College of Medicine, Case Western Reserve University, Cleveland, OH 44195, USA

* To whom correspondence should be addressed:
Daniel Blankenberg[1,2], Computational Life Sciences, Cleveland Clinic Research, Cleveland Clinic, 9500 Euclid Avenue, NA2, Cleveland, OH 44195, USA
Email: blanked2@ccf.org

† These authors contributed equally to this work.

**Background** – Classical approaches to protein structure prediction are largely limited by molecular dynamics sampling timescales and force-field accuracy, the combinatorial conformational search required by ab initio methods, or a strong reliance on evolutionary data in the case of deep learning methods. Current quantum approaches to protein structure prediction (QPSP) remain constrained by substantial qubit, connectivity, circuit-depth, and measurement requirements, which grow at rates that depend strongly on the molecular representation and encoding and currently preclude realistic off-lattice simulations of complex proteins on near-term hardware.

**Methods** – We propose and implement a logarithmic-scale variational quantum eigensolver that reduces the qubit requirement for (N) torsional degrees of freedom to $O\left(log_2 N\right)$. In statevector simulations, molecular torsions are derived from gauge-fixed relative phases of the complex statevector amplitudes. On quantum hardware, where these phases are not directly accessible through computational basis measurement, a distinct measurement compatible decoder maps the empirical cumulative distribution function (CDF) derived from measured basis-state probabilities to bounded torsional variables. Both decoders feed the same downstream pipeline, in which a classical reconstruction algorithm builds full heavy-atom coordinates from the decoded backbone and side-chain torsions. The architecture employs an EfficientSU2 ansatz for conformational generation and classical physics-based objective functions for structural evaluation. The primary objective function is a customizable hybrid quantum-classical Hamiltonian comprising many conventional molecular mechanics force field terms, including Lennard-Jones dispersion and attractive forces, Coulombic electrostatics, implicit solvent effects via a sigmoid-weighted Solvent Accessible Surface Area (SASA) approximation, explicit hydrogen bonding, as well as local and non-local steric and clash potentials. As a benchmark, alternate objective functions were also incorporated, including the Rosetta all-atom energy function and the OpenMM AMBER-based energy function. The optimization landscape is traversed using a two to three stage relaxation strategy, progressing from mechanical collapse to physical refinement and entropic relaxation, to mitigate the barren plateau problem common in high-dimensional VQE landscapes. The output heavy-atom models were then protonated, parameterized, and minimized further using GROMACS with the amber99sb-ildn force field.

**Results** – We evaluated QTF on two standard miniature protein benchmarks, chignolin (5AWL; 10 residues, 166 atoms) and Trp-cage (2JOF; 20 residues, 284 atoms), using a customizable empirical energy function, Rosetta, and OpenMM. For each protein, up to 400 independent replicas were performed per energy function, producing up to 1,200 replica trajectories per target. Each replica retained up to 5,000 low-energy conformational snapshots, allowing trajectory-level analysis beyond the final optimization endpoint. For chignolin, QTF sampled highly native-like conformations, with the best retained snapshot reaching 0.623 Å Cα RMSD and the best final model reaching 1.199 Å. Trp-cage was more challenging, with no sub-2.0 Å structures observed, but snapshot mining improved the best RMSD from 3.512 Å among final models to 2.501 Å among retained snapshots. Across all three energy functions, low-energy structures did not always correspond to the lowest-RMSD conformations, highlighting persistent energy-ranking imbalance in the sampled landscapes. The custom energy function produced the strongest final-model behavior overall, including lower RMSDs and structural metrics closer to experiment, while all three backends benefited from mining retained low-energy snapshots. Finally, saved circuit parameters from selected chignolin models were executed in 300 jobs each on ibm_cleveland and ibm_miami. Native-like structures were recovered in 88/300 *ibm_cleveland* jobs and 45/300 *ibm_miami* jobs, with best hardware RMSDs of 1.758 Å and 1.782 Å, respectively. These results demonstrate recovery of native-like conformations across two QPU architectures, although the lower recovery rate and broader structural distributions on *ibm_miami* indicate substantial backend-dependent variability.

**Conclusions** – This work demonstrates that high-resolution, off-lattice protein structure prediction is feasible with exponentially fewer qubits than traditional lattice models. By converting physical qubit constraints into circuit depth constraints, the framework architecture establishes a scalable foundation for hybrid quantum biophysics pipelines on near-term hardware. Major limiting factors include a high sensitivity to the choice of energy function, and significant computational overhead associated with hardware-compatible measurement and readout, together with the additional cost that would be required for direct relative-phase recovery. Nonetheless, our work provides to the best of our knowledge the first quantum algorithm for protein structure prediction in all-atom continuous space, laying the groundwork for the nascent QPSP field to begin moving away from coarse grain Cα lattices and onto more realistic representations of proteins in nature.

## INTRODUCTION

### The Protein Folding Problem

Protein structure prediction (PSP) remains a challenging problem in biological research. Lying at the intersection of physics, chemistry, and biology, it remains one of the most computational demanding tasks, in part due to the combinatorial nature of the exponentially growing conformational space as the protein sequence length increases. Machine learning (ML) and artificial intelligence (AI) based methods, including the deep learning approaches such as AlphaFold [1,2] and RosettaFold [3–5] (along with their growing number of derivative methods), have made significant advancements in this field, helping researchers generate highly accurate models by leveraging the wide breadth of experimentally determined proteins now available in public repositories such as the Protein Data Bank (PDB) [6]. The need for more physics-based PSP methods still remains, in part due to the fact that the rate at which novel protein sequences are being discovered is far outpacing the rate at which the corresponding structures are being experimentally determined [7]. In other words, there is a growing amount of structural biodiversity that still has yet to be uncovered, presenting a challenge for methods relying heavily on known structures for training data.

Classically, physics-based methods have their own challenges. Whether the method is based on molecular dynamics simulations, where native-like structures are attempted to be found by simulating from initial unfolded decoy structures in hopes that the force field drives trajectories to sampling the folded state's thermodynamic basin, or classical *ab initio* structure prediction approaches which are driven by iterative conformer enumeration and scoring, it is still fundamentally an optimization of a massive combinatorial search space. The emergence of quantum algorithms and their potential to tackle such problems by mapping solutions into the Hilbert Space presents a paradigm where *ab initio* PSP could become more scalable. Consequently, the field of protein structure prediction with quantum computing (herein referred to as QPSP) has gained traction in recent years [7–20]. In parallel, advanced methods of scoring known conformers (without necessarily predicting the structures themselves) [21–25], as well as estimation of binding free energies in multi-scale modeling [26,27] by applying first-principles quantum chemistry approaches on quantum computers, is rapidly evolving as well.

### The Quantum Bottleneck

Although the QPSP algorithms and workflows mentioned above have paved the way for much of this field, they present several limitations inherently affecting the ability to model proteins in a more realistic representation. The vast majority of these methods employ coarse grain lattice models, where space is discretized on a lattice and each node usually corresponds to the amino acid's alpha carbons (Cα). This representation is commonly known as "alpha trace" and the scoring mechanism that helps distinguish between unfolded and folded structures usually involves the use of statistical or knowledge-based potentials derived from pairing frequencies between amino acids in a database of experimentally determined structures. The two most common scoring functions are the Miyazawa-Jernigan (MJ) potentials [28–31] and the more basic hydrophobic-polar (HP) model [30,31]. The former is more explicit, in which unique contact potentials are assigned for each possible amino acid contact pair (from the canonical 20 amino acid set) and scaled according to known pairing frequencies, while the latter is a more rudimentary alphabet, where residues are either classified as either hydrophobic or polar (with only hydrophobic-hydrophobic pairs being scored favorably). Dating back to the 1980s and 1990s, this more fundamental form of modeling proteins was still useful in terms of capturing important properties such as hydrophobic collapse, density, solvent accessible surface area (SASA), and other more global effects in protein folding. Decades later, this reduced representation became the norm in the early days of QPSP as a practical means to handle the resource constraints in near-term quantum hardware. In these quantum algorithms, a classical hamiltonian is usually generated for a protein sequence, where qubits are used for encoding both the turns or positions on the lattice ("the lattice walk") and the nearest-neighbor interactions between occupied lattice nodes. Thus, a discrete alpha trace lattice model would require significantly less qubits and potentially circuit depth when compared to all-atom models in a more continuous space. Nonetheless, these reduced models still lack the ingredients to produce realistic proteins and model their interactions with other proteins and ligands, which is crucial in drug design and biological research at-large. Another limitation is that many of the these methods require encoding constraint terms (such as overlap constraints which help prevent any two amino acids from occupying the same lattice node) which can lead to rather massive Hamiltonians even for small peptides, and enforcing these constraints becomes challenging in today's quantum optimization approaches [10].

### The New Quantum Frontier: Modeling in Continuous Angular Space

As hardware capabilities improve, so must the algorithms. While the lattice models have been foundational, the future of QPSP depends on modeling methods that are equal parts realistic and scalable. In this work, we present a novel algorithm encoding protein torsion (backbone and sidechain) by default, and optionally backbone 3-body angles, into qubit phases. The method, Quantum Torsion Folder (QTF), improves on state-of-the-art quantum methods by: 1) moving away from discretized space and towards a more continuous representation, 2) modeling all heavy-atoms with full side chains, 3) incorporating sophisticated molecular mechanics energy functions that capture interactions far beyond the alpha trace level, 4) logarithmic scaling with respect to the number of protein angles and the corresponding qubits.

## MATERIALS AND METHODS

The methodology presented herein introduces a fundamental shift in how quantum circuits parameterize molecular geometries. Rather than mapping discrete spatial coordinates to individual qubits, which inherently forces models onto artificial spatial lattices, we propose a logarithmic encoding strategy that maps continuous internal degrees of freedom directly into the phase amplitudes of a quantum state vector. To implement this theory, we formulated a hybrid quantum-classical pipeline utilizing a generator-evaluator paradigm. A parameterized quantum circuit functions as the conformational generator exploring the continuous, non-convex conformational space, while a classical physics-based objective function serves as the evaluator to assess structural viability. Figure 1 below illustrates the key steps incorporated in this method.

The following sections detail the mathematical encoding strategy, the differentiable geometric reconstruction, the formulation of the classical energy landscape, and the multi-stage variational optimization curriculum required to navigate it.

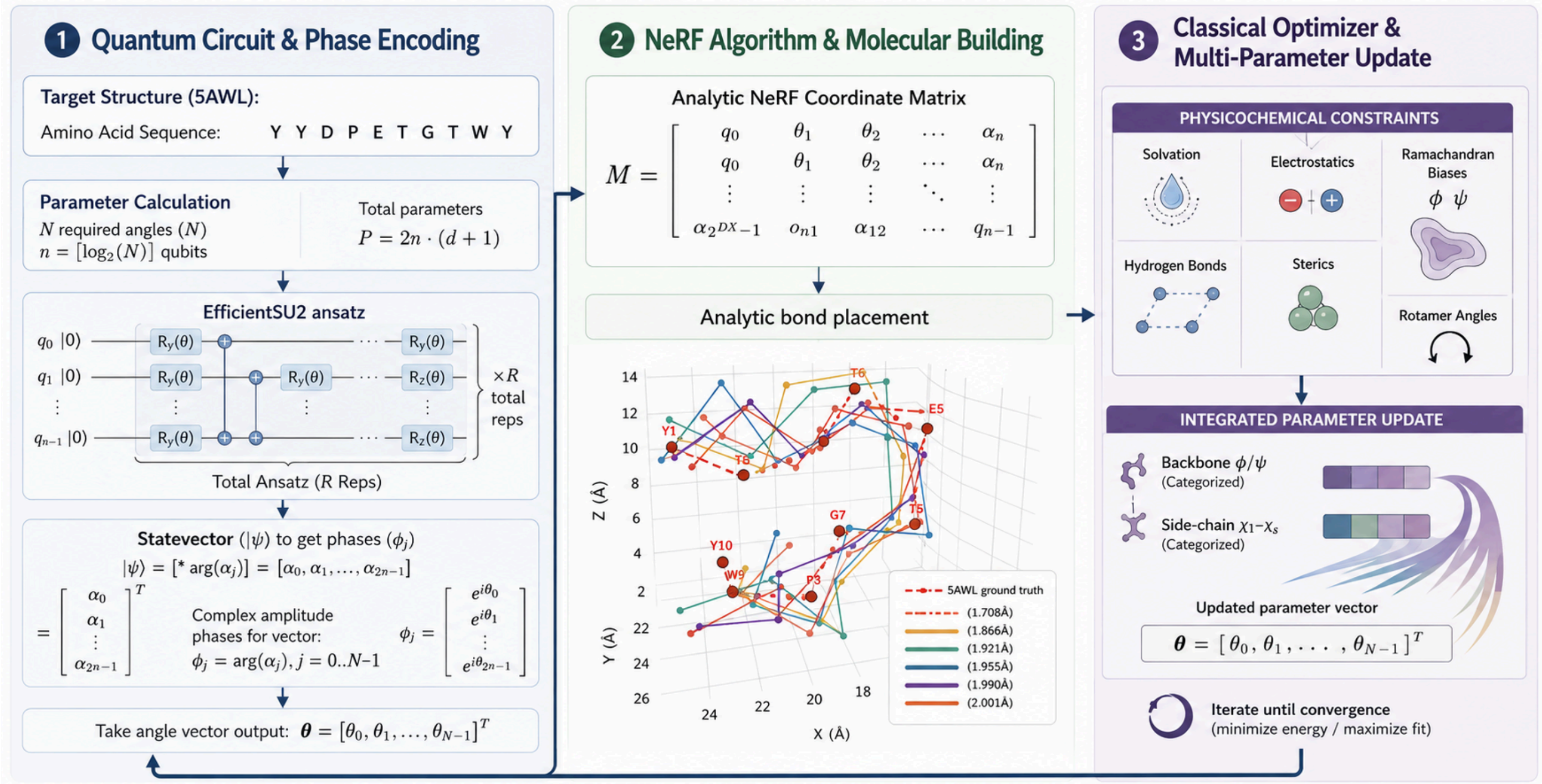

**Figure 1:** Schematic overview of the hybrid quantum-classical continuous-space protein structure prediction pipeline. The closed-loop VQE architecture consists of four interconnected phases. (I) Quantum State Generation: an n-qubit parameterized quantum circuit, scaled logarithmically relative to the torsional degrees of freedom, generates a complex statevector. The gauge-fixed relative phases of the resulting amplitudes are extracted to directly represent continuous molecular torsion angles in the $[-\pi, \pi)$ domain; (II) Differentiable Coordinate Reconstruction: the continuous 1D vector of torsion angles is passed through the deterministic NERF algorithm. This step analytically rebuilds the full heavy-atom 3D Cartesian coordinates of the polypeptide chain using idealized bond lengths and angles, preserving end-to-end differentiability; (III) Physical Energy Evaluation: the reconstructed 3D topology is scored by a classical physics-based objective function that computes a total scalar energy based on structural viability; (IV) Variational Optimization Curriculum: the total scalar energy is fed back into classical optimizers to update the quantum circuit parameters.

### Methodological Pipeline Overview

The proposed approach is structured as a closed-loop variational quantum eigensolver (VQE) that seamlessly maps quantum parameter spaces to classical 3D cartesian coordinates. The methodology is executed through four primary, interconnected phases:

1. Quantum State Generation: an $n$-qubit parameterized quantum circuit, scaled logarithmically such that $n = \lceil log_2 N \rceil$ for $N$ degrees of freedom, is prepared to produce a complex statevector. The gauge-fixed relative phases of the resulting complex amplitudes are extracted and mapped to bounded, continuous torsion angles in the domain $[-\pi, \pi)$. These angles mathematically represent the backbone torsions $(\phi, \psi, \omega)$, side-chain rotamer $(\chi_1 - \chi_4)$, and optionally 3-body backbone angle terms $(\tau, \Theta)$ degrees of freedom.

Since the peptide-bond ω torsion is known to predominantly occupy the trans region near 180 degrees, QTF supports multiple treatments for this degree of freedom. The ω torsion can be excluded from the circuit and fixed at trans, included as an optimizable DOF constrained to a 170-190 degree trans-like window, or included as a fully variable torsion over the full angular range from -π to π. In the windowed mode, the circuit parameter is mapped into the 170-190 degree interval, avoiding wasted optimizer effort on highly unfavorable cis-like ω values.

2. Differentiable Coordinate Reconstruction: the extracted torsion angles are processed through a deterministic Natural Extension Reference Frame (NERF) algorithm [32,33] by default, with other optional reconstructors also made available. Operating purely in angular space with fixed bond lengths and angles derived from standard crystallographic parameters, this step analytically reconstructs the 3D cartesian coordinates of all heavy atoms ($N$, $C\alpha$, $C$, $O$ backbone atoms, and full side chains heavy atoms) without iterative spatial refinement. For many amino acids where torsions are naturally fixed due to a planar geometry (such as those with aromatic rings) or involve symmetry in the final torsions (such as lysine, arginine, glutamate, and aspartate), the sidechain is capped with these end groups using a template library, while preserving rotation freedom in the $\chi_N$ torsions leading up to them. For a detailed look at QTF's side chain torsional degrees of freedom, see Figure S2.

3. Physical Energy Evaluation: the reconstructed all heavy-atom 3D topology is scored by a composite, physics-based objective function. This evaluator computes a total scalar energy by aggregating distinct physical attractive and repulsive terms, including implicit solvation (hydrophobic effect), explicit hydrogen bonding, Coulombic electrostatics, softened Lennard-Jones steric repulsion, disulfite bridging, end empirical geometric constraints (e.g., Ramachandran distributions) and clash penalties.

4. Variational Optimization Curriculum: the total scalar energy is fed back into gradient-free and gradient-based classical optimization algorithms (COBYLA and SLSQP). These optimizers calculate the necessary updates to the quantum circuit parameters (θ), closing the hybrid loop and continuously minimizing the conformational energies calculated from the resulting angle vector. To mitigate the barren plateau problem inherent in high-dimensional quantum optimization, this step is executed iteratively across a three-stage curriculum designed to guide the target system from initial hydrophobic collapse to refined entropic relaxation.

**Encoding and Quantum Ansatz**

Traditional quantum algorithms for chemistry and biology often employ a direct spatial mapping, one or more qubits are assigned to represent a discrete degree of freedom such as a lattice site, a relative lattice turn, or an atomic coordinate.

An n-qubit register spans a $2^n$ dimensional complex Hilbert space. A normalised pure state in this space is written in the computational basis as

$$|\psi(\theta)\rangle = \sum_{j=0}^{2^n-1} r_j e^{i\phi_j}|j\rangle \text{ , where } n = \lceil log_2 N\rceil \text{ and N = degrees of freedom.}$$

where θ is the vector of trainable circuit parameters, $r_j \geq 0$ is the real-valued amplitude magnitude, and $\phi_j \in (-\pi, \pi]$ is the principle complex phase of the j-th basis-state amplitude.

The normalization condition $\sum_{j=0}^{2^n-1} r_j 2 = 1$ constrains the amplitude magnitudes. However, a global phase is physically unobservable, so a statevector with ($2^n$) indexed amplitudes provides at most ($2^{n-1}$) independent relative phases. The phases reachable in practice are further restricted by the parameterized circuit ansatz. QTF therefore uses gauge-fixed relative phases, rather than absolute amplitude phases, as molecular torsion variables. These relative phases are wrapped into the interval $[-\pi, \pi)$ as defined below.

Setting the circuit width to $n = \lceil log_2 N\rceil$ ensures that the state vector carries at least *N* indexed amplitude slots, whose gauge-fixed relative phases provide *N-1* independent phase coordinates. We extract the first (N) indexed amplitudes and compute their gauge-fixed relative phases and use them directly as the continuous backbone ($\phi, \psi, \omega$) and side-chain ($\chi_1$, …, $\chi 4$) dihedral angles required for full-atom reconstruction. The reference component is fixed to zero and assigned to the undefined N-terminal ($\phi$) slot, while the remaining (N-1) relative phases parameterize the physical torsional degrees of freedom. For proteins with up to N ≈ 1,024 degrees of freedom, only n = 10 qubits are needed. This strategy reduces the qubit overhead to $O(\lceil log_2 N\rceil)$, achieving an exponential compression relative to direct encoding, a reduction that places the register size firmly within the coherence and connectivity limits of near-term quantum hardware.

A crucial conceptual distinction separates two different angle-valued quantities in this framework.

**Circuit Parameters** $\theta$

The vector $\theta \in \mathbb{R}^P$ contains the arguments of the parameterised rotation gates (e.g., $Rz(\theta_k)$ and $Ry(\theta_k)$) inside the ansatz circuit. These are real numbers that control how the quantum state is prepared; they are the latent variables optimised by the classical minimiser. They are not themselves molecular torsion angles and carry no direct geometric meaning.

**Torsion Angles** $\phi_j$

In statevector simulation, molecular torsion angles are derived from the gauge-fixed relative phases of the complex statevector amplitudes. For a parameterized quantum state

$$|\psi(\theta)\rangle = \sum_j \psi_j(\theta)|j\rangle \text{ ,} \qquad \psi_j(\theta) = \langle j|\psi(\theta)\rangle$$

QTF defines the phase-derived torsion variable associated with basis-state index j as

$$\tilde{\phi}_j(\theta) = wrap_{[-\pi, \pi)}(arg(\langle j|\psi(\theta)\rangle) - arg(\langle 0|\psi(\theta)\rangle))$$

Equivalently,

$$\tilde{\phi}_j(\theta) = Arg_{[-\pi,\pi)}(\psi_j(\theta)\psi_0^*(\theta))$$

Here, arg(z) denotes the phase of a complex number z, defined modulo 2π. The operator *wrap*[−π, π) shifts an angle by an integer multiple of 2π to select its representative in the interval [−π, π). Correspondingly, Arg[−π, π)(z) denotes the principal representative of the phase of z on that branch. The asterisk denotes complex conjugation, and $\psi_0$ is the amplitude of the reference basis state |0…0⟩.

A quantum state is physically defined only up to a global phase: |ψ⟩ and $e^{(i\gamma)|\psi\rangle}$ represent the same physical state. Under this transformation, every amplitude phase is shifted by the same value γ. Subtracting the phase of the reference amplitude cancels this common shift, ensuring that the decoded torsion vector (and therefore the reconstructed protein conformation) is invariant under global phase.

This gauge convention fixes $\tilde{\varphi}_0 = 0$, leaving N − 1 independent relative phases among N indexed amplitudes. In the implemented ordering of molecular degrees of freedom, the reference slot corresponds to the N-terminal φ angle. This dihedral is undefined because the first residue has no preceding peptide bond and does not affect coordinate reconstruction. The remaining relative phases are mapped to the allowed physical ranges of the corresponding backbone and side-chain torsions.

The circuit parameters θ remain latent optimization variables and are not themselves molecular torsion angles. The classical optimizer adjusts θ, while the relative-phase decoder maps the resulting statevector to the torsion variables used for three-dimensional coordinate reconstruction and energy evaluation. This relative-phase decoder is used in the exact statevector simulation workflow. The separate probability-CDF decoder used for hardware execution is described below and does not measure or reconstruct these statevector phases.

**Overflow of the Quantum process**

The quantum component of this framework is realised through a parameterised quantum circuit whose architecture must satisfy two simultaneous requirements: it must be expressive enough to encode as much of the full conformational diversity of the target protein as possible, and shallow enough to remain executable within the gate fidelity and coherence constraints of near-term hardware. The choice of circuit architecture therefore directly determines the quality of the encoded geometry, the cost of each energy evaluation, and the practical feasibility of execution on a physical quantum processor.

To generate the parameterized quantum state, we use Qiskit's hardware-efficient EfficientSU2 ansatz [34] by default. For $N_{torsion}$ encoded torsional degrees of freedom, the number of qubits is chosen as $n = \max(2, \mathrm{ceil}(\log_2 N_{torsion}))$, and the number of repetition layers is set heuristically to $d = \mathrm{ceil}(N_{torsion}/n) + 2$. Each repetition contains independent single-qubit Ry and Rz rotations on every qubit, controlled by the trainable parameter vector θ, followed by a circular nearest-neighbor CX (CNOT) entangling layer. The ansatz also includes a final Ry/Rz rotation block after the last entangling layer, giving P = 2n(d + 1) trainable parameters. Thus, the parameter count grows linearly with both register width and circuit depth, while the register size grows logarithmically with the number of encoded torsions. As a concrete example, a 6-qubit circuit with five repetition layers contains 72 trainable parameters acting on a 64-dimensional Hilbert space.

In addition to the default EfficientSU2 ansatz, QTF also supports an optional brickwork ansatz. In this circuit, each repetition applies Ry/Rz rotations followed by two nearest-neighbor CX layers: one over even-indexed qubit pairs and one over odd-indexed qubit pairs. This produces local nearest-neighbor entanglement with reduced serial entangling depth relative to circular entanglement, making it a useful alternative for hardware-oriented experiments.

In a classical simulation environment, the full statevector is directly accessible, allowing torsion angles to be extracted deterministically by reading the phase of each amplitude relative to the reference amplitude. This direct phase extraction pathway is used throughout the optimization loop, where thousands of energy evaluations are performed and exact phase information ensures rapid, reliable convergence.

On physical quantum hardware, however, the statevector is not directly observable. Because standard computational-basis measurements yield probabilities $P(x) = \left|\alpha_x\right|^2$ that are inherently stripped of complex phase information, recovering the exact simulator phases would require costly interference measurements. Instead, for hardware execution, we utilize a divergent readout strategy: the angle information is mapped directly from the empirical probability distribution, which is obtained by running the circuit for a fixed number of shots.

This hardware-compatible mapping proceeds by computing the cumulative distribution function (CDF) of the measured probability vector and mapping it linearly onto the torsion range [−π,π], where the two endpoints represent the same angular direction. Given the measured computational-basis probabilities ($P_j$), the cumulative probability associated with basis-state index (i) is

$$F_i = \sum_{j=0}^{i} P_j, \qquad F_i \in [0, 1]$$

The corresponding hardware-decoded variable is then defined as

$$\tau_i^{CDF} = 2\pi F_i - \pi$$

Consequently, $\tau_i^{CDF} \in [-\pi, \pi]$, with the two endpoints representing the same angular direction. The first (N) values of the resulting angle vector are used as the molecular degrees of freedom for full heavy-atom reconstruction.

This CDF mapping is monotone, preserves the full shape of the underlying amplitude distribution, and guarantees that every extracted angle falls within the physically valid torsion range without any additional clipping or wrapping. The two decoders feed the same downstream coordinate construction and scoring pipeline, but they are mathematically distinct circuit-to-torsion mappings and need not produce the same conformation from identical circuit parameters. In particular, the empirical CDF mapping does not reconstruct or approximate the relative phases used by the statevector decoder. Accordingly, the calculations from the hardware experiments constitute a parameter-transfer experiment using a measurement-compatible probability decoder, rather than hardware recovery of the simulator-derived torsions. The circuit parameters are transferred between the workflows, but their conversion into molecular geometry differs.

Executing this methodology on physical Quantum Processing Units (QPU) inherently requires this transition from exact statevectors to measurable observables. While full quantum state tomography could theoretically recover the exact encoded phases, it scales exponentially and is prohibitive for large registers. The CDF approach requires only computational-basis measurements and retains the logarithmic circuit width scaling. Its shot complexity, however, depends on the number of populated basis states and the probability and angular resolution required for structural reconstruction. Furthermore, future work could investigate whether classical-shadow or expectation-value-based measurements provide useful alternative hardware-compatible observables for torsion decoding or optimization.

**Parameterization of Degrees of Freedom and Coordinate Reconstruction**

To maintain computational tractability while preserving biological realism, our method models the polypeptide as a continuous kinematic chain. High-frequency, stiff localized vibrations, such as bond stretching and bond-angle bending, are generally decoupled from the quantum optimization problem. Instead, covalent bond lengths and internal bond angles are held constant at ideal crystallographic values derived from the Engh and Huber parameters [35,36]. However, we once again note that these parameters can also be included as degrees of freedom (DOFs) if desired, largely by QTF's incorporation of PHEAT, which permits a wide range of customizability in the encoded search space and runtime conditions exposed in user-defined recipes. As mentioned previously, the peptide bond dihedral ($\omega$) can be firmly constrained to standard planar geometries ($\pi$ rad for trans configurations, 0 rad for cis-proline), but can also be sampled fully or within an allowed window.

By isolating these parameters and focusing on the more flexible backbone and side-chain torsions, the quantum generator focuses exclusively on the highly flexible, low-frequency degrees of freedom that govern the global fold. For a protein of length $L$, the target parameters are strictly reduced to the continuous torsion space:

- Backbone Dihedrals: the rotational angles $\phi(N - C\alpha - C - N)$ and $\psi(C\alpha - C - N - C\alpha)$ for each residue, and the peptide bond torsional angles $\omega(C\alpha - C - N - C\alpha)$ when included as a DOF.
- Side-Chain Rotamers: up to four $\chi$ angles ($\chi_1$ through $\chi_4$) per residue, strictly determined by the amino acid topology (e.g., zero for Glycine and Alanine; four for Arginine and Lysine).

Thus, the vector of gauge-fixed relative phases extracted from the quantum statevector directly constitutes the complete conformational angle vector $\Theta = \left[\phi_0, \psi_0, \chi_{1,0}, ..., \phi_{L-1}, \psi_{L-1}, ...\right]$.

While the quantum generator operates natively in relative-phase torsion space, the classical evaluator (the physical energy function) relies on Euclidean distances to compute spatial interactions, such as steric clashes, hydrogen bonding, Coulombic electrostatics, and Lennard-Jones interactions. Consequently, a mathematically exact, computationally efficient bridge is required to project the 1D torsion vector $\Theta$ into a full heavy-atom 3D Cartesian geometry.

To achieve this without resorting to computationally expensive iterative spatial solvers, the pipeline employs the Natural Extension Reference Frame (NERF) algorithm [32,33] by default, or optionally an internal-coordinate/Z-matrix atom placement rebuilder imported from the PHEAT library. NERF provides a deterministic, analytical transformation from internal coordinates to global Cartesian space. Given the known 3D coordinates of three preceding atoms, $A$, $B$, and $C$, alongside a standard bond length $l_{CD}$, standard bond angle $\theta_{BCD}$, and the quantum-proposed torsion angle $\tau_{ABCD} \in \Theta$, the exact Cartesian coordinate of the fourth atom $D$ is computed as:

1. Construct a localized, orthonormal reference frame based on the unit vector $BC$, the normal to the plane defined by $A$, $B$, and $C$, and their cross product.

2. Define the position of atom $D$ within this local reference frame using standard trigonometric projections. The local coordinates $(x, y, z)$ are calculated directly as $x = l\,cos(\pi - \theta)$, $y = l\,cos(\tau)\,sin(\pi - \theta)$, and $z = l\,sin(\tau)\,sin(\pi - \theta)$.

3. Project these local coordinates into the global Cartesian reference frame using the rotational matrix derived from the localized orthonormal vectors, and translate the result relative to the position of atom $C$.

This projection is applied sequentially along the kinematic chain. The algorithm first traces the backbone $\left(N_i \rightarrow C\alpha_i \rightarrow C_i \rightarrow N_{i+1}\right)$ while appropriately branching to place the carbonyl oxygens. Upon establishing the backbone scaffold, the procedure recurses down the side-chain topologies, placing heavy atoms ($C\beta$, $C\gamma$, etc.) based on the corresponding $\chi$ angles.

Crucially, the entire NERF projection is mathematically smooth and fully differentiable. This ensures that spatial gradients calculated by the classical evaluator in 3D space can be effectively backpropagated through the geometrical reconstruction and mapped to the quantum circuit parameters, enabling efficient navigation of the VQE landscape.

**The Classical Hamiltonian**

In quantum chemistry, the Hamiltonian is the energy operator in the molecular Schrödinger equation; solving this equation yields molecular energy eigenvalues for a given electronic structure model, while variational methods minimize the expectation value of this operator. Here, we use the term classical Hamiltonian to refer to an empirical, potential energy function that evaluates the macroscopic stability of a quantum-generated protein conformation. Unlike electronic structure VQE, the total scalar score $E_{total}$ is not decomposed into quantum observables and measured as the expectation value of a qubit Hamiltonian; it is computed classically from the reconstructed

Cartesian structure after the circuit readout and minimized by the classical optimizer. There are several available objective functions that can be used at runtime. For optional comparison to established scoring methods, we incorporate the Rosetta centroid and all-atom energy functions using PyRosetta [37,38], and OpenMM using the AMBER ff14SB based scorer [39,40]. Additionally, PHEAT can generate empirical statistical contact potentials inspired by DFIRE-like distance-dependent scoring [41]. This allows users to derive potentials from curated PDB structure sets, with resolutions ranging from coarse alpha-carbon or residue-level contacts, analogous to Miyazawa-Jernigan contact potentials [28,42], to more explicit heavy-atom or full-atom type specificity. All of these scoring mechanisms can be imported into QTF via the PHEAT library.

Much of our efforts entailed developing a new customizable energy function, with terms inspired by scorers and molecular mechanics force fields known to be influential in both promoting physically valid structures, as well as helping discriminate between decoy and native-like conformers. This custom objective function is a linear combination of distinct physical and empirically defined energies and penalties:

$$E_{total} = E_{constraint} + E_{elec} + E_{hbond} + E_{burial} + E_{rama} + E_{rotamer} + E_{geom} + E_{omega} + E_{\pi-stack} + E_{steric} + E_{disulfide}$$

To evaluate the reconstructed 3D Cartesian coordinates, atoms are assigned coarse physicochemical parameters, including effective partial charges and van der Waals radii, partly inspired by AMBER-family molecular mechanics force fields. Because early circuit-generated conformations can contain severe atomic overlaps, direct use of an unmodified Lennard-Jones 12-6 potential can produce extremely large energies that destabilize the classical optimizer. To improve numerical stability, we define a softened steric contribution, $E_{steric}$, that accounts for both local adjacent-residue contacts and longer-range nonbonded interactions. This steric contribution is decomposed into explicit clash penalties, $E_{clash}$, and van der Waals-like packing interactions, $E_{VdW}$.

The clash term penalizes physically impossible intra- and inter-residue heavy-atom overlaps, including local adjacent-residue clashes that may not be fully captured by longer-range nonbonded masks. $E_{VdW}$ contains atom type-specific repulsive and attractive components analogous to conventional Lennard-Jones interactions. The attractive component weakly promotes favorable nonbonded packing contacts, whereas the repulsive component provides a softened excluded-volume penalty based on van der Waals radii. For extreme overlaps, the repulsive penalty is logarithmically capped, for example $E = 50 + log(E - 49)$, preventing highly compressed initial conformations from numerically dominating the objective while preserving an informative optimization signal.

Electrostatic and hydrogen bonding contributions are evaluated as pairwise/geometric interaction terms. Electrostatics are modeled using a Coulomb-like potential over $E_{elec}$ assigned effective partial charges, with a distance floor to prevent numerical divergence at very short separations. Hydrogen bonds are modeled using a geometric distance-and-angle potential. For each backbone nitrogen, a virtual amide hydrogen is constructed from the local peptide geometry. Carbonyl oxygen acceptors from non-neighboring residues are rewarded when the H-O distance is less than 3.5 A and the N-H-O geometry satisfies a directional linearity criterion, providing a structural bias toward plausible backbone hydrogen-bonding patterns.

Because explicitly modeling solvent molecules would introduce many additional degrees of freedom, QTF instead uses an implicit hydrophobic burial proxy, $E_{burial}$. For hydrophobic atoms, a soft neighbor-counting function approximates local burial from nearby atomic contacts. Hydrophobic atoms with low burial fractions receive an exposure penalty, biasing the search toward compact conformations with improved hydrophobic packing. An additional cysteine-specific term is included, where plausible sulfur-sulfur contacts compatible with disulfide formation are rewarded.

To constrain the vast search space and enforce biological realism, the Hamiltonian further incorporates empirical structural biases:

- Ramachandran and Rotamer Wells ($E_{rama} + E_{omega} + E_{rotamer}$): Gaussian energy wells are strategically placed in the continuous $(\phi, \psi)$ space corresponding to the highly populated $\alpha$-helix and $\beta$-sheet regions, alongside strong penalties for sterically forbidden zones. Similar harmonic and Gaussian potentials guide side-chain $\chi_1$ angles toward crystallographically observed rotamer states (e.g., *gauche+*, *trans*). Additionally, when ω is treated as an optimizable degree of freedom, $E_{omega}$ is introduced. This term is decomposed into $E_{omega-center}$, which mildly favors trans peptide geometry near 180 degrees, and $E_{omega-window}$, which penalizes raw ω proposals outside the allowed trans window.

- Geometric Integrity ($E_{geom}$): hard penalties enforce physical invariants that are blind to the torsion-space generator, such as strict L-chirality at the $C\alpha$ chiral center, peptide plane twist minimization, and Proline ring-closure distance constraints.

By formulating the objective function as a mostly continuous, numerically guarded landscape, the classical scorer can evaluate highly compressed early conformations while providing a stable optimization signal that guides the search toward physically plausible structures. For a detailed mathematical description of the full potential and each of these terms, please see the supplementary material.

**Multi-Stage Optimization Curriculum**

The energy landscape of a polypeptide chain is famously characterized by a rugged, highly non-convex topography containing multitudes of local minima (Levinthal's paradox). In the context of a Variational Quantum Eigensolver (VQE), this challenge is exacerbated by the initialization phase: a randomly parameterized quantum statevector generally corresponds to a highly extended, sterically chaotic, and unphysical conformation. A naive application of a single optimization algorithm across this landscape inevitably leads to rapid entrapment in local minima or stalls due to vanishing gradients (the barren plateau problem).

To systematically guide the quantum generator from a random initialization to a physically viable thermodynamic minimum, the methodology employs a dynamic, multi-stage optimization curriculum. Three stages are run when the custom energy function is employed, while two stages are used with any of the other energy functions. With the custom energy function, this approach mimics a simulated annealing strategy by progressively modulating the strength of artificial physical constraints, specifically the end-to-end hairpin bias ($\lambda$) and the hydrophobic surface tension ($\gamma$), while strategically alternating between derivative-free and gradient-based classical solvers. It is important to note that the third stage, where these two parameters are tapered, only applies when the custom energy function is used. For both the Rosetta and OpenMM energy functions, which lack these parameters, the optimization is limited to the first two stages accordingly.

Stage 1: Global Hydrophobic Collapse – The primary objective of the initial stage is to force the extended, pseudo-random polypeptide chain into a compact, globular topology. To achieve this, the classical evaluator applies maximal constraint weights ($\lambda = 50.0$, $\gamma = 15.0$), creating a steep, artificially deep energy funnel toward collapsed states.

Crucially, the optimization in this stage is driven by COBYLA (Constrained Optimization BY Linear Approximations) [43,44]. At this early stage, the energy landscape is incredibly noisy. Minor adjustments to backbone torsions can cause wild fluctuations in steric penalties. Gradient-based methods are inherently greedy and would immediately follow these noisy local slopes into shallow traps (e.g., locking into a single, isolated hydrogen bond while the rest of the chain remains extended). COBYLA, as a derivative-free optimizer, ignores these misleading local gradients. By taking broader steps based on linear approximations of the objective function, it effectively jumps over local topological noise, successfully driving the macroscopic hydrophobic collapse and locating the correct general basin on the energy landscape.

Step 2: Local Structural Refinement – Once COBYLA has driven the chain toward a rough compact state, the broad collapse objective has largely been established. However, this collapsed state can still contain severe local geometric conflicts, including overlapping van der Waals radii, unfavorable packing contacts, distorted torsional preferences, and suboptimal hydrogen-bond geometries.

At this stage, the optimization strategy switches to SLSQP (Sequential Least-Squares Programming) to perform local refinement of the circuit parameters. Rather than searching broadly across conformational space, SLSQP refines the continuous torsion representation within the compact basin produced by the initial collapse stage. Although the objective includes softened and guarded terms rather than a strictly smooth analytic force field, the continuous torsional parameterization allows numerical gradient-based optimization to make finer adjustments than the derivative-free COBYLA stage. This refinement can reduce steric clashes, improve packing, and favor more compatible hydrogen-bond and torsional geometries, while remaining subject to the nonconvexity and approximate nature of the empirical scoring function.

Step 3: Natural Relaxation – In the final custom-scoring stage, the collapse-driving bias terms are relaxed relative to the first two stages: the hydrophobic burial coefficient is reduced from $\gamma = 15.0$ to $\gamma = 2.5$, and the end-to-end constraint strength is reduced from $\lambda = 8.0$ to $\lambda = 1.5$. This transition removes the strong biasing funnel and allows the true physical potentials, namely electrostatics, hydrogen bonding, and rotamer preferences, to dominate the objective function. SLSQP is maintained as the optimizer to guide the final descent. This relaxation phase allows the protein to expand slightly, setting into a more stable thermodynamic minimum without being artificially compressed, finalizing the hybrid quantum-classical folding process.

**Ensemble Folding and Statistical Ranking**

Due to the vast and highly degenerative nature of the phase-torsion landscape, a single VQE trajectory is statistically unlikely to locate the global thermodynamic minimum. Different initial parameterizations of the quantum generator will inevitably collapse into distinct local basins. To ensure robust and reproducible conformational sampling, we employ an ensemble-based folding strategy governed by deterministic initialization.

To guarantee perfect reproducibility across computational environments, the random number generator seed for each ensemble is explicitly derived from the cryptographic SHA-256 hash of the input amino acid sequence. Individual replicas within the ensemble are subsequently initialized using unique offset seeds. For a multi-replica ensemble, replicas are run serially, with the random seed hash changing based on each replica index. For larger parallel runs across an HPC, QTF can be executed for single replicas in each task, with the seed then changing based on the slurm array task id. This ensures that every job is initialized uniquely. Before starting the multi-stage optimization curriculum, each replica executes a broad basin-hopping scout protocol. A predetermined number of random quantum parameter vectors ($\theta$) are evaluated by the classical Hamiltonian, based on the corresponding torsion vector. The vector yielding the lowest initial potential energy is selected as the starting coordinate for that specific replica. This ensures that while each replica starts in a geometrically diverse region of the phase space, it begins in a localized basin favorable to rapid hydrophobic collapse, significantly reducing the computational waste of optimizing highly improbable, high-energy extended chains. Upon completion of the optimization curriculum across all independent replicas, the methodology must distinguish between a globally converged fold and a rugged landscape trapping replicas in divergent local minima.

Finally, the methodology isolates two distinct primary metrics for structural ranking: the lowest thermodynamic energy and the highest structural accuracy (when an experimental X-ray or NMR ground truth is available). Our framework explicitly decouples these metrics. The lowest energy structure represents the absolute minimum found on the respective energy function landscape, serving as the primary prediction metric for sequences without a known ground truth 3D structure. Conversely, the lowest-RMSD structure represents the prediction closest to the experimental ground truth. While a perfectly calibrated, generalizable energy function may help reduce the gap between lowest-energy and lowest-RMSD models, empirical approximations in the Hamiltonian often decouple them, leading to imbalances in the landscape. To this day and to the best of our knowledge, there is no truly generalizable force field/scoring function for *ab initio* PSP. It is a challenging problem on its own, and an entire separate field is dedicated to their ongoing development. Fundamentally, there is a significant difference in the reliability between using a force field in molecular dynamics simulations that start with a ground state structure and instead using it to fold a protein where the trajectory or optimization path begins from a largely unfolded state. For one, in the first scenario, the structure is usually already in a global minimum state, particularly in the case of monomeric structures in solution, so the force field-driven trajectory allows one to sample the conformational ensemble within the low-energy basin the structure is already in. Folding a protein starting from random, unfolded, high-energy states using the same force field is an entirely different task, with high-energy barriers often preventing a direct traversal to the ground-truth structure using conventional approaches. Enhanced sampling methods, such as replica exchange and metadynamics approaches do provide some advantage, but it is still challenging and the best results are usually demonstrated within a specific subset of proteins and not necessarily proving universality of the scoring method. Thus, in QTF we provide a custom energy function that can be calibrated and further optimized by the user, as well as state-of-the-art established scorers like Rosetta and OpenMM. By reporting both the energy rank and the RMSD rank independently alongside physics-based metrics (e.g., radius of gyration, end-to-end distance), the methodology provides a transparent assessment of both the quantum optimization's efficiency and the energy function's biological fidelity. Optionally, all structures in the ensemble can be further optimized using GROMACS [45], where explicit hydrogens are added to the structures and further relaxed via a steepest descent minimization using a chosen force field (amber99sb-ildn by default [46]).

## RESULTS

### Quantum simulation of continuous torsional space successfully samples native-like conformations

We chose to test this approach with two of the commonly benchmarked proteins in structure prediction methods: the 166 atoms, 10 residue chignolin (PDB: 5AWL [47,48]) and the 284 atom, 20 residue Trp-cage mini protein (PDB: 2JOF [49,50]). Chignolin is a small peptide characterized as a compact antiparallel beta-hairpin, largely stabilized by aromatic packing between Tyrosine and Tryptophan residues. On the other hand, Trp-cage is a larger designed miniprotein that folds into a compact helix-containing structure, with an N-terminal alpha helix packed against a C-terminal polyproline-like segment around a buried tryptophan-centered hydrophobic core. It exhibits a helix, turn and loop all in one structure, making it an ideal candidate for validating a PSP method's ability to sample and predict proteins with more diverse structural features.

We performed 1,200 replica simulations for each protein, with 400 replicas assigned to each energy function. This was performed on an HPC cluster, with each replica run as its own SLURM array task, and initialized from random circuit parameters using the previously mentioned scouting procedure. By default, 50 random parameter sets were evaluated, and the parameter set producing the lowest initial cost-function value was selected as the starting point for full optimization. To ensure reproducibility and variability, unique seeds were used in each replica initialization, based on the protein sequence and replica task ID. Varying the starting point across replicas allowed the workflow to sample a broader region of torsional space rather than repeatedly optimizing from the same basin. The vast majority of these jobs completed, with the exception of approximately 67 jobs using the OpenMM energy function, which failed at the 2nd optimization stage due to numerical instability in the optimizer in those trajectories.

For each successful replica, the workflow attempted to retain up to 5,000 low-energy snapshot structures from the optimization trajectory, in addition to the final selected model. These snapshots corresponded to the lowest-energy structures encountered during optimization according to the raw backend energy function. To reduce structural redundancy among very similar low-energy candidates, retained snapshots were filtered by requiring them to differ by at least 0.25 raw energy units from previously retained structures. This energy spacing threshold was applied in backend-specific units: kcal/mol for the custom energy function, Rosetta energy units (REU) for Rosetta, and kJ/mol for OpenMM. Therefore, each replica retained up to 5,000 snapshots extracted from the final optimization stage, but not necessarily exactly 5,000.

| Protein | Final Replica Models | Retained Snapshots | Total Structures |
| --- | --- | --- | --- |
| 2JOF | 1,176 | 4,650,631 | 4,651,807 |
| 5AWL | 1,157 | 2,681,469 | 2,682,626 |
| **Total** | 2,333 | 7,332,100 | 7,334,433 |

**Table 1:** Accounting of successful final replica models and retained snapshots for both proteins across all three energy functions.

This snapshot-retention strategy was designed to expose imbalances in the energy-function landscapes. The final evaluated model from an optimization trajectory is not guaranteed to be the structure closest to the experimental ground truth. Retaining low-energy snapshots therefore provides a way to identify native-like structures that were visited during optimization but not necessarily selected as the final endpoint. This is also consistent with the physical view that protein folding does not necessarily yield a single static structure, but rather an ensemble of thermodynamically related conformations near the global minimum basin, as commonly observed in NMR structural ensembles.

Across the current selected analysis set, 2,333 final models and 7,332,100 retained snapshots were analyzed for a total of 7,334,433 protein structures across 2JOF and 5AWL (see Table 1). The presented analyses, including folding funnels, RMSD distributions, native-like hit rates, snapshot-versus-final comparisons, energy-RMSD correlations, and lowest-RMSD structure collections were computed from these retained models.

| Protein | Residues | Torsions | Bins/torsion | Search space | Fraction sampled by 5,000 |
| --- | --- | --- | --- | --- | --- |
| Chignolin (5AWL) | 10 | 44 | 3 | $9.85 \times 10^{20}$ | $5.08 \times 10^{-18}$ |
| Chignolin (5AWL) | 10 | 44 | 6 | $1.73 \times 10^{34}$ | $2.89 \times 10^{-31}$ |
| Chignolin (5AWL) | 10 | 44 | 12 | $3.05 \times 10^{47}$ | $1.64 \times 10^{-44}$ |
| Trp-cage (2JOF) | 20 | 88 | 3 | $9.70 \times 10^{41}$ | $5.16 \times 10^{-39}$ |
| Trp-cage (2JOF) | 20 | 88 | 6 | $3.00 \times 10^{68}$ | $1.67 \times 10^{-65}$ |
| Trp-cage (2JOF) | 20 | 88 | 12 | $9.29 \times 10^{94}$ | $5.38 \times 10^{-92}$ |

**Table 2:** Estimated conformational space for chignolin and Trp-cage under coarse angular discretizations. Search space size was calculated as $b^d$, where $b$ is the number of bins per torsion and $d$ is the number of QTF torsional degrees of freedom.

To provide scale for the conformational search problem, we estimated the size of a hypothetical discretized torsional space using coarse angular binning. This discretization is used only as an illustrative reference; QTF does not sample torsions in fixed bins, but instead optimizes continuous torsional degrees of freedom. As seen in Table 2, even under an extremely coarse discretization of only three bins per torsion, the resulting space is approximately $10^{21}$ states for chignolin and $10^{42}$ states for Trp-cage. The retained snapshots from each replica are therefore not arbitrary samples from the full conformational landscape, but the lowest energy structures encountered during continuous final optimization stages. Native-like conformations recovered within this retained snapshot pool indicate that the optimization trajectory reached low energy basin states, even when the final selected replica model (the last evaluated model in the VQE optimization) was not itself the lowest-RMSD structure.

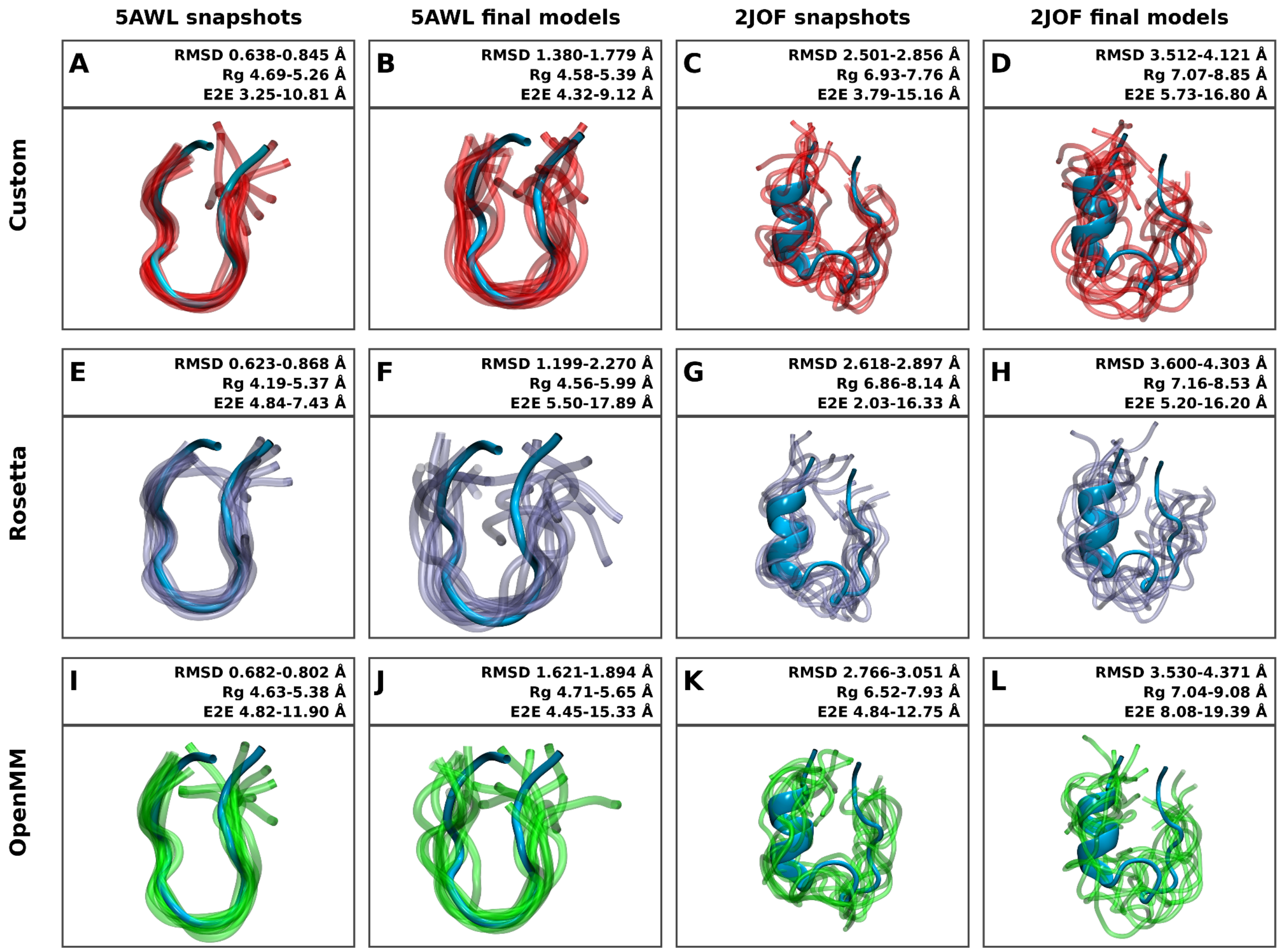

**Figure 2:** Structural overlays of the 10 lowest RMSD snapshots and 10 final models for each protein, model class, and energy function. Rows correspond to energy functions: custom (A-D), Rosetta (E-H), and OpenMM (I-L). Columns show 5AWL snapshots, 5AWL final models, 2JOF snapshots, and 2JOF final models, respectively. Each panel reports the RMSD range for the 10 displayed structures. Predicted structures are colored by energy function, with experimental reference structures shown in cyan.

The structural overlays in Figure 2 show the 10 most accurate snapshots and final replica models, as sorted by RMSDs. The image demonstrates a coherent narrative where near-native 5AWL beta-hairpin conformations are found across all three energy functions, whereas the corresponding final model overlays are broader and often displaced from the experimental structure. This visual comparison is consistent with the quantitative snapshot mining results: native-like or near-native conformations can be present among the low-energy retained snapshots even when the final selected replica model is not the lowest-RMSD structure. For 2JOF, both snapshot and final model overlays remain more heterogeneous and farther from the experimental Trp-cage fold, matching the absence of sub-2.0 Å structures in the numerical analysis. In both cases, this observation indicates that the snapshot ensemble could be representative of the low energy basin of the landscape, where many native-like structures co-exist.

| **Protein** | 5AWL | 5AWL | 5AWL | 2JOF | 2JOF | 2JOF |
|---|---|---|---|---|---|---|
| **Energy** | Custom | Rosetta | OpenMM | Custom | Rosetta | OpenMM |
| **Final RMSD med.** | 2.90 | 4.20 | 3.84 | 5.39 | 6.59 | 6.21 |
| **Final RMSD best** | 1.38 | 1.20 | 1.62 | 3.51 | 3.60 | 3.53 |
| **Snapshot RMSD med.** | 1.22 | 1.28 | 1.36 | 3.24 | 3.32 | 3.49 |
| **Snapshot RMSD best** | 0.64 | 0.62 | 0.68 | 2.50 | 2.62 | 2.77 |
| **Exp. E2E** | 5.51 | 5.51 | 5.51 | 11.25 | 11.25 | 11.25 |
| **Final E2E med.** | 7.84 | 16.78 | 14.86 | 14.11 | 24.66 | 21.61 |
| **Best Snapshot E2E med.** | 7.36 | 7.88 | 7.85 | 10.09 | 9.97 | 9.89 |
| **Exp. Rg** | 4.85 | 4.85 | 4.85 | 6.91 | 6.91 | 6.91 |
| **Final Rg med.** | 5.16 | 6.57 | 6.22 | 7.91 | 10.05 | 9.15 |
| **Best Snapshot Rg med.** | 5.02 | 5.04 | 5.03 | 7.46 | 7.48 | 7.50 |

**Table 3:** Structural accuracy and compactness of final replica models and best retained snapshots. Median and best RMSD values are reported for final replica models and for the per-replica best retained snapshots, defined as the lowest-effective-RMSD snapshot from each replica. End-to-end distance (E2E) and radius of gyration (Rg) are reported as medians for the experimental reference, final replica models, and the same per-replica best snapshots used for the snapshot RMSD statistics. Values are reported in angstroms.

This table separates final-model quality from the best structures visited during each optimization trajectory. Across both proteins, the per-replica best snapshots consistently have lower median and best RMSD values than the final replica models, indicating that many trajectories sampled more native-like conformations than were ultimately selected as endpoints.

A notable pattern is that the custom energy function produces final models with substantially more native-like compactness than Rosetta or OpenMM. For both 5AWL and 2JOF, the median final-model end-to-end distance (E2E) and radius of gyration (Rg) are much closer to the experimental values under the custom energy function. This suggests that, even when final RMSD remains imperfect, the custom scoring function better preserves global chain dimensions and overall compactness in the selected endpoint models. In contrast, Rosetta and OpenMM final models are more expanded, particularly in end-to-end distance, indicating that their endpoint selection tends to favor conformations with less native-like global geometry.

For 5AWL, the best-snapshot RMSD medians are near 1.2-1.4 Å across all three energy functions, with the absolute best snapshots below 0.7 Å, effectively recovering the experimental structure. These values show that native-like beta-hairpin conformations were frequently present in the retained snapshot pools. The best-snapshot Rg values are also close to the experimental Rg, suggesting that these low-RMSD snapshots recover the compactness of the native structure more faithfully than the final models. The best-snapshot E2E values remain somewhat larger than the experimental value, indicating residual variation in terminal geometry even among the most native-like snapshots.

For 2JOF, snapshot mining also improves RMSD relative to final models, but the best-snapshot medians remain above 3 Å and the absolute best observed snapshots remain above the 2 Å native-like threshold. The best-snapshot E2E and Rg values are closer to the experimental structure than the final-model values, especially for Rosetta and OpenMM, but the remaining RMSD gap indicates that compactness alone is insufficient to recover the full Trp-cage fold.

Overall, the table supports two complementary conclusions: the custom energy function better maintains native-like global compactness in final endpoint models, while snapshot mining reveals that all three energy functions can visit substantially improved structures during optimization. Thus, model selection remains a key bottleneck, but the custom energy function appears to provide a more favorable baseline for preserving experimentally realistic chain dimensions.

**Trajectory snapshot mining systematically recovers higher-accuracy structures than final model selection**

Because ground-state protein structures exist as ensembles of low-energy conformations rather than a single static geometry, relying exclusively on the final minimized endpoint of a variational trajectory may inadvertently overlook more native-like intermediate structures. To investigate this, we compared the structural accuracy of the final selected model from each of the 2,333 completed replica runs against the lowest-RMSD conformation retained within its corresponding 5,000-snapshot pool.

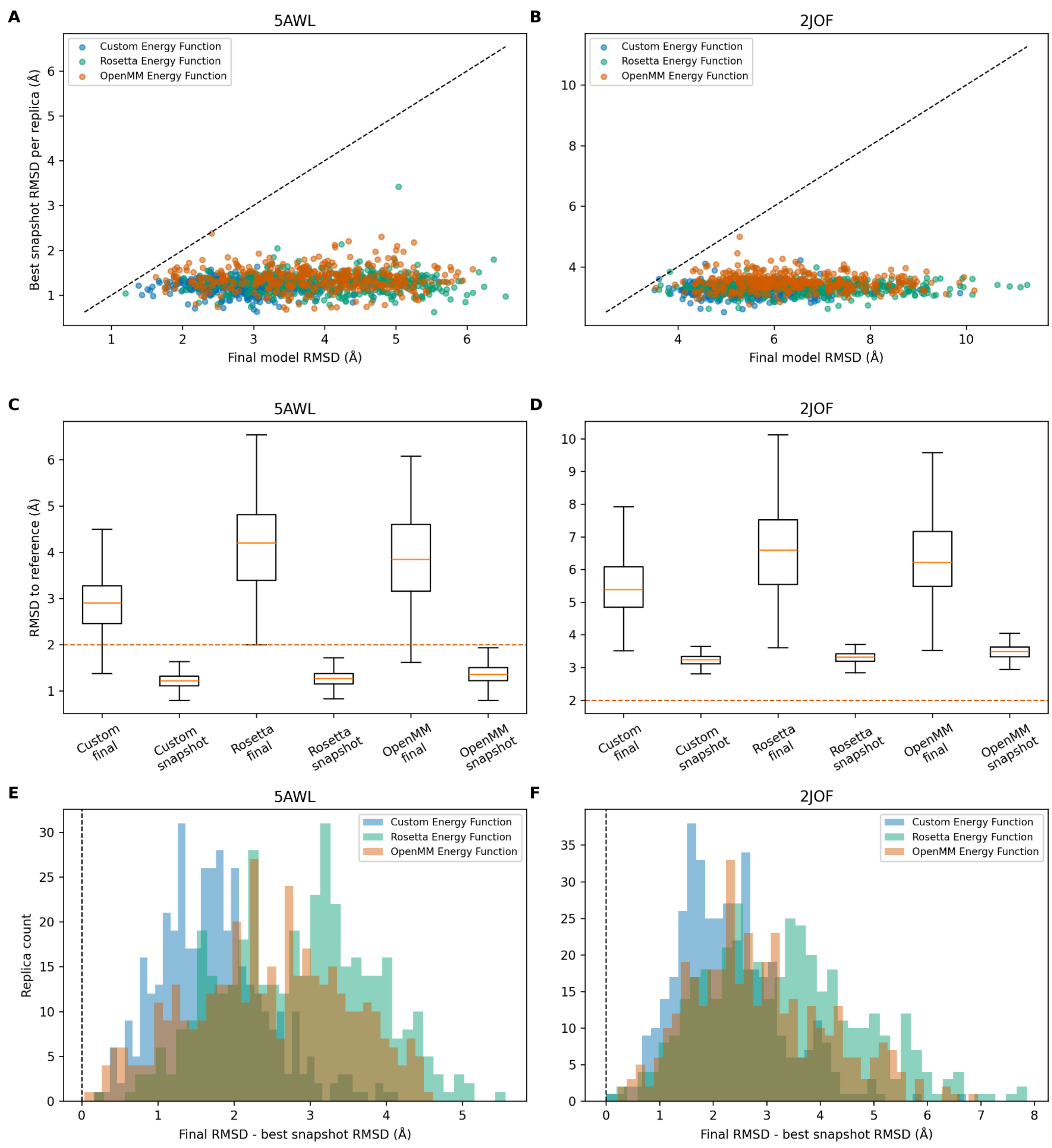

**Figure 3:** Comparison final replica models with the best retained snapshot from the same replica. Panels A and B plot final-model RMSD versus best-snapshot RMSD for 5AWL and 2JOF, respectively; points below the diagonal indicate replicas where snapshot mining recovered a lower-RMSD structure than the final selected model. Panels C and D summarize the corresponding RMSD distributions for final models and per-replica best snapshots across each energy function, with the 2.0 Å horizontal line marking the native-like threshold. Panels E and F show the per-replica RMSD improvement, defined as final-model RMSD minus best-snapshot RMSD, for 5AWL and 2JOF. In all snapshot panels, the snapshot value is the single retained structure with the lowest effective RMSD from that replica, not the distribution of all saved snapshots.

Figure 3A and 3B compare each replica's final model RMSD against the lowest-RMSD retained snapshot from the same replica. Points below the diagonal indicate that the trajectory sampled a better structure than the final selected model. 5AWL snapshots cluster around ~1-1.5 Å RMSD across all three energy functions, while final models span a broader range. This shows that many 5AWL replicas sampled near-native structures even when the final model was less accurate, indicating that model selection is a major bottleneck. 2JOF on the other hand shows the same qualitative pattern but at higher RMSD. Best snapshots are mostly around ~3-4 Å, while final models extend much farther, often above 6 Å. Thus, snapshot mining improves 2JOF substantially, but the sampled structures generally remain short of the 2.0 Å native-like threshold.

Figure 3C and 3D illustrate that the best-snapshot RMSD distributions are distinctly narrower and possess significantly lower medians than the final-model distributions. Because each best-snapshot value represents an extreme statistic (i.e., the minimum RMSD observed over thousands of retained conformations) this best-of-trajectory extraction successfully isolates the underlying accuracy of the sampling algorithm before it is subject to final-model selection biases.

The divergence between structures sampled and the structures ultimately selected is most apparent in the native-like hit rates for chignolin (5AWL). Snapshot mining achieved a near-perfect recovery rate: 100% of custom-energy replicas, 98% of OpenMM replicas, and 99.3% of Rosetta replicas contained at least one snapshot breaking the 2.0 Å native-like threshold. In stark contrast, relying solely on the final selected models yielded native-like hit rates of only 6%, 4%, and 0.3%, respectively. Furthermore, the absolute highest-accuracy conformation found in 5AWL reached an RMSD of 0.623 Å within the Rosetta snapshot pool, whereas the best final selected model across any backend was only 1.199 Å.

While the larger, more topologically complex Trp-cage (2JOF) target presented a significantly harder sampling challenge with no final model or sampled snapshot crossing the 2.0 Å threshold, snapshot mining still yielded crucial improvements. The best observed RMSD improved from 3.512 Å (final model) to 2.501 Å (snapshot pool). Additionally, the median RMSD improvement gained via snapshot mining was 2.18 Å for the custom energy function, 2.66 Å for OpenMM, and 3.26 for Rosetta, demonstrating that the trajectories consistently navigated toward the native basin before diverging.

To directly quantify this selection gap on a per-replica basis, Figure 3E and 3F visualizes the distribution of RMSD improvement, defined as the final-model RMSD minus the lowest snapshot RMSD. The heavily positive skew across all panels definitively separates the success of the quantum conformational sampling from the limitations of final-model selection. The positive values indicate that the optimization trajectory frequently visited and evaluated a more native-like conformation than the endpoint structure it ultimately selected. Therefore, the logarithmic quantum generator is highly capable of sampling the correct conformational space: the primary bottleneck preventing a higher yield of native-like final models lies in the classical evaluation and endpoint selection mechanism.

### Objective function landscapes reveal endpoint selection is limited by imperfect energy funneling

To understand the divergence between the structural accuracy of the broadly sampled snapshot ensembles and the final selected endpoint models, we evaluated the folding funnel landscapes generated during optimization. The raw-energy funnels reveal that each backend imposes a distinct internal structure on the sampled conformational space. For example, the raw custom energy landscape (Figure 4, Panel A) heavily concentrates native-like 5AWL snapshots into a distinct low-RMSD region, yet it still fails to cleanly rank these highest-accuracy structures as the lowest-energy structures among the sampled snapshots. Conversely, Rosetta and OpenMM exhibit broad low-energy bands that encompass both near-native and much higher-RMSD conformations, presenting a relatively flat selection gradient at the bottom of the optimization basin.

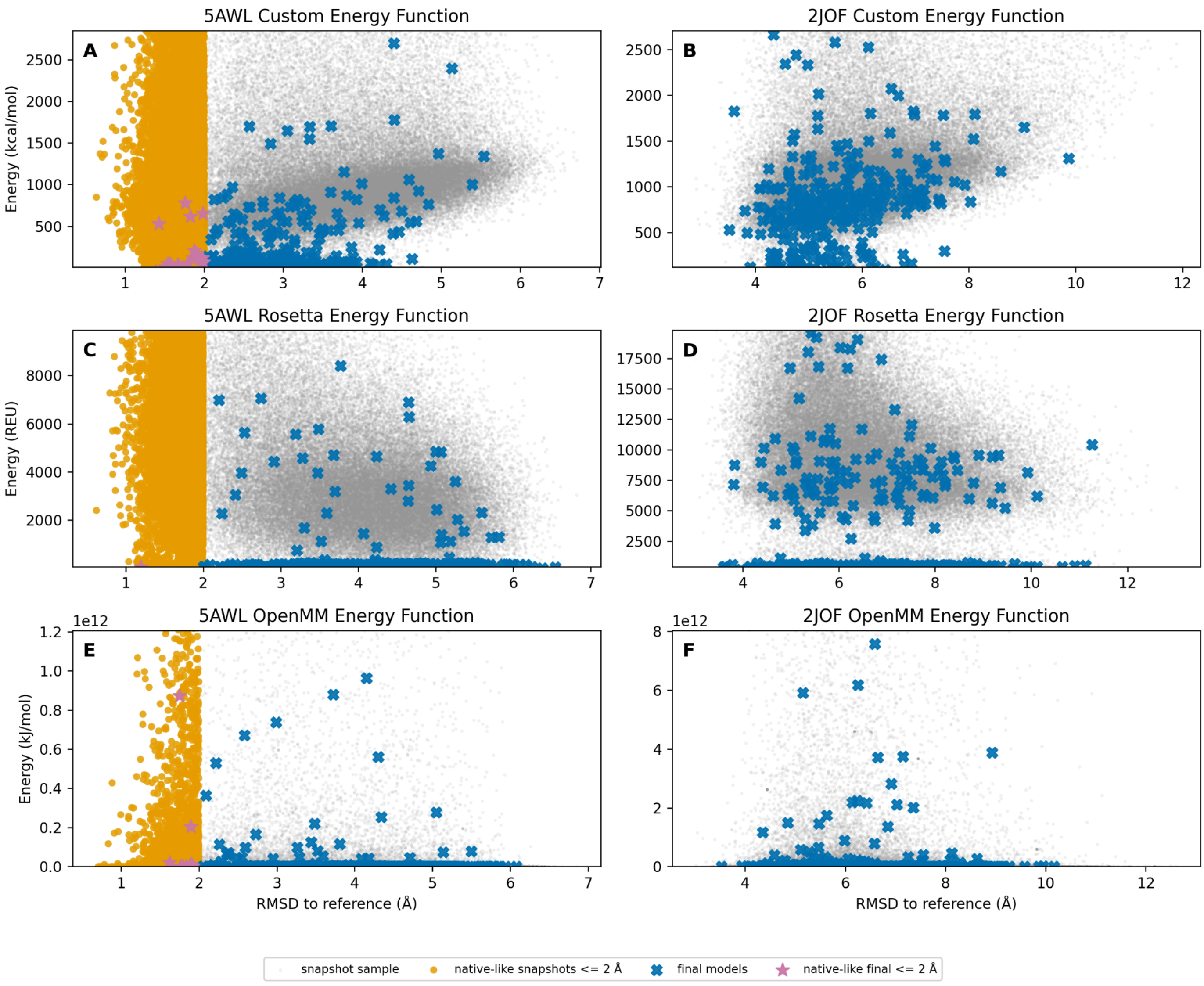

**Figure 4:** Raw-energy folding-funnel landscapes. The raw energy units are kcal/mol for the custom energy function, Rosetta energy units (REU) for Rosetta, and kJ/mol for OpenMM. These panels test whether each backend's own scoring objective produces an internal funnel toward lower RMSD, but raw y-values should not be compared quantitatively across energy functions because the scales and definitions differ.

Because these raw energies are reported in backend-specific units and scales, the panels should be interpreted within each protein-energy-function combination rather than compared directly across energy functions. The key observation is therefore not the absolute energy value, but whether lower raw energy corresponds to lower RMSD within a given backend. The weak separation between the lowest-energy and lowest-RMSD structures suggests that optimization often reaches regions containing native-like conformations, but the scoring objectives do not consistently provide enough resolution near the basin floor to select those conformations as final endpoints.

To further statistically quantify these landscape observations, we computed the Spearman rank correlations between structural accuracy (RMSD) and the calculated energies of the final selected models (Figure 5). In these plots, a positive Spearman correlation indicates that larger energy-term values are associated with higher RMSD and therefore poorer structural accuracy, whereas a negative correlation indicates that larger term values are associated with lower RMSD, or equivalently that lower term values may favor less native-like structures. Correlations near zero indicate weak monotonic coupling between that term and structural accuracy.

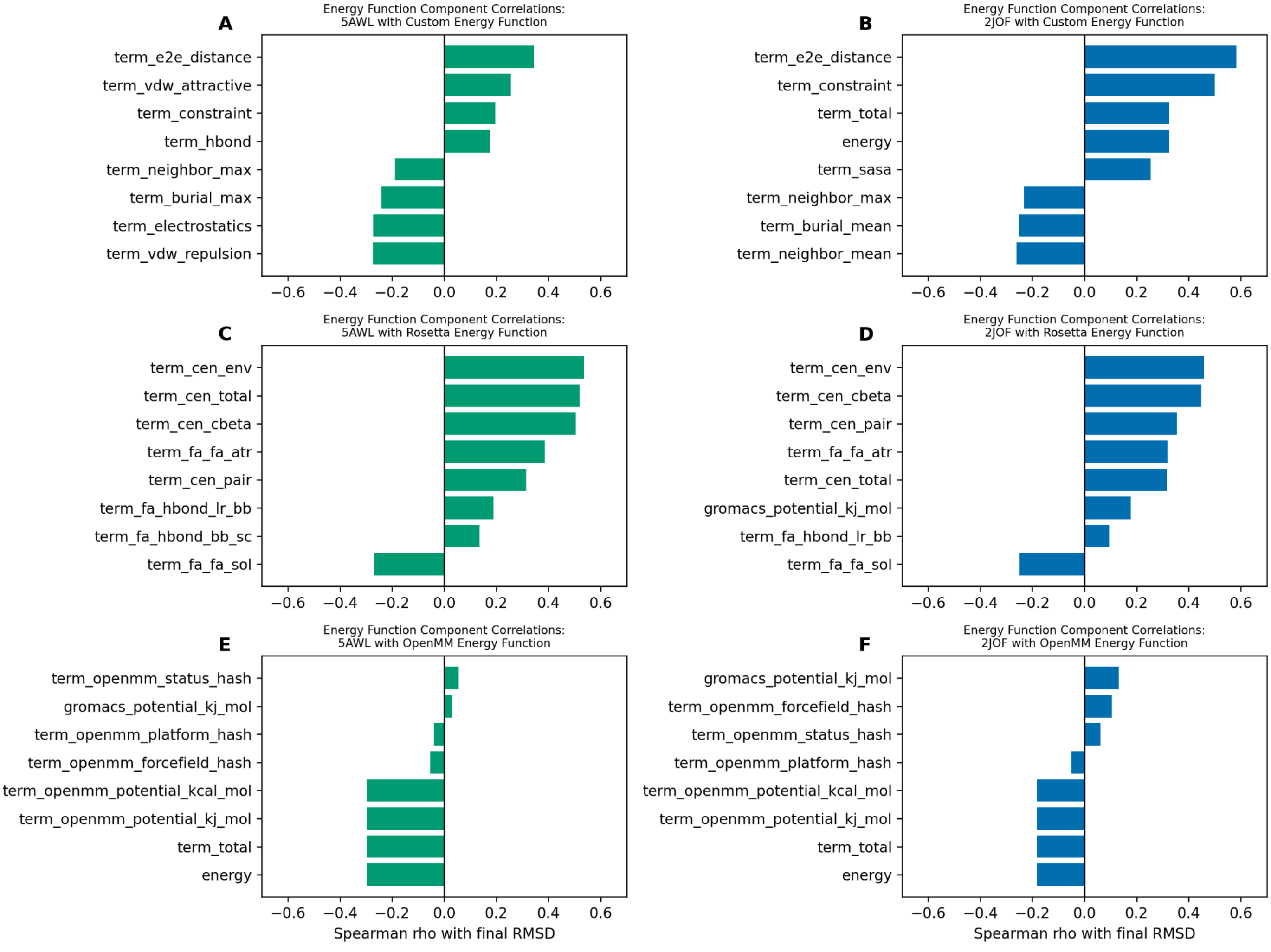

**Figure 5:** Final-mode energy-term correlations. The strongest final-model energy-term Spearman correlations with RMSD for each protein and energy function. Panels follow the same row-major layout as

Figure 1: (A) 5AWL custom, (B) 2JOF custom, (C) 5AWL Rosetta, (D) 2JOF Rosetta, (E) 5AWL OpenMM, and (F) 2JOF OpenMM. These panels are intended to identify which terms most strongly track structural accuracy among final selected models.

The final-model energy-term correlations suggest that the three scoring backends differ substantially in how their internal terms relate to structural accuracy. The custom energy function shows a more balanced mixture of positive and negative correlations, indicating that no single class of terms dominates the relationship with RMSD. Instead, structural accuracy appears to emerge from competing contributions, including compactness, end-to-end distance, burial, neighbor contacts, sterics, and electrostatics. This balance may help explain why the custom energy function samples native-like 5AWL structures relatively effectively, but it also suggests that the current weighting does not yet provide a clean monotonic ranking of the best structures.

Rosetta shows a more penalty-like pattern, especially for 5AWL, where many of the strongest centroid and full-atom terms are positively correlated with RMSD. This implies that increases in Rosetta energy terms often track worsening structural accuracy, which is the expected direction for an energy function used for selection. However, the folding-funnel plots (Figure 4) show that Rosetta can still produce broad low-energy regions containing both near-native and higher-RMSD structures, so these correlations are useful but not sufficient for unambiguous final-model selection.

OpenMM shows the opposite tendency: total OpenMM potential energy is negatively correlated with RMSD in both proteins. This means that lower OpenMM energies are not consistently associated with lower RMSD final models in this dataset, and may even favor structures that are physically favorable under the force field but topologically farther from the experimental fold. Thus, OpenMM minimization provides physically meaningful relaxation, but its raw potential energy alone appears poorly aligned with native-like model selection for these QTF-generated conformations.

Together, these correlations reinforce the main selection problem: the simulations often visit low-RMSD conformations, but the scoring terms do not yet provide a uniformly reliable ordering of those structures. The custom energy function appears more balanced, Rosetta provides stronger penalty-like signals, and OpenMM appears dominated by physical stabilization terms that do not necessarily encode fold-level native similarity.

Finally, it is worth noting that navigating a continuous, off-lattice optimization space introduces unique physical challenges. Visual inspection of several high-quality, low-RMSD snapshots revealed instances of physically suspect GROMACS energies (following post-processing of the heavy-atom models) driven by unconstrained conformational artifacts, such as inferred terminal bond formation occurring when the N- and C-termini were placed in abnormally close covalent proximity. While the quantum algorithm's logarithmic phase encoding excel at rapidly sampling the vast continuous space, these edge-case artifacts highlight the necessity for downstream structural review, including automated filtering based on terminal separation, non-local heavy-atom contacts, and ring penetrations, prior to final model selection.

**Executions on quantum hardware preserve native-like predictions**

To further validate QTF, we performed 300 executions on quantum hardware using *ibm_cleveland*, a 156-qubit IBM Heron R2 QPU with heavy-hex topology, and another 300 using *ibm_miami*, a 120-qubit IBM Nighthawk R1 QPU with square topology. The transpiled circuits used native single-qubit RZ, SX, and X gates, with CZ gates serving as the two-qubit entangling operation. These experiments were conducted with a “warm start” strategy, whereby the circuit was initialized and sampled using previously optimized parameters from the simulations. These parameters were chosen from the top 10 replica final models shown in Figure 2, for each energy function. Since the chignolin (5AWL) simulations appeared to yield substantially more native and native-like structures than Trp-cage (2JOF), the former was selected for these quantum hardware experiments. This 30 replica simulation ensemble presented final structures with RMSDs ranging from 1.199 to 2.270 Å, with a median of 1.784 Å. Since these structures were derived from the last cost function evaluation at the end of the multi-stage optimization, these final circuit parameters would present ideal starting conditions for the subsequent jobs executed on the QPU. To account for stochasticity in transpilation layout, qubit assignment, and device calibration, each of the 30 initial circuits was executed 10 times, yielding 300 quantum hardware jobs per QPU per protein, for a total of 600. This repeat strategy reduced dependence on any single compiled circuit instance, since one execution could map onto qubits with higher error rates or less favorable calibration states, whereas another could map onto a more favorable physical layout. Each job comprised a single execution and sampling of the parameter-bound circuit using the SamplerV2 primitive with 8,192 shots, rather than continued iterative optimization. Error-suppression and randomized-compilation techniques were applied using 32 twirled circuit randomizations, XY4 dynamical decoupling, and both gate and measurement twirling.

| Metric | Range (ibm_cleveland) | Range (ibm_miami) | Median (ibm_cleveland) | Median (ibm_miami) |
|---|---|---|---|---|
| Logical QTF qubits | 6 | 6 | 6 | 6 |
| Circuit depth | 443-506 | 206 | 455 | 206 |
| Total gates | 975-1067 | 449-450 | 1048 | 449 |
| 1Q gates | 807-860 | 389-390 | 842 | 389 |
| 2Q gates | 168-208 | 60 | 204 | 60 |
| 2Q depth | 143-160 | 60 | 152 | 60 |
| sx operations | 454-507 | 192 | 493 | 192 |
| rz operations | 332-373 | 197-198 | 349 | 197 |
| cz operations | 168-208 | 60 | 204 | 60 |

**Table 4:** Transpilation details on executed quantum hardware circuits. Ranges and medians are reported across all 300 jobs for each QPU.

The transpiled circuits were generally consistent across the 300 executions on each hardware backend. All jobs used the same QTF circuit involving six logical qubits, reflecting the fixed torsional encoding for 5AWL. Variation in the compiled circuits therefore arose from backend-aware transpilation, physical qubit layout selection, routing, and calibration conditions rather than changes in the logical circuit size or underlying EfficientSU2 ansatz.

On *ibm_cleveland*, circuit depth ranged from 443 to 506, with a median of 455, while the total gate count ranged from 975 to 1067, with a median of 1048. The two-qubit gate burden was also relatively stable: CZ counts ranged from 168 to 208, with a median of 204, and two-qubit depth ranged from 143 to 160, with a median of 152. The single-qubit gate composition was dominated by SX and RZ operations. SX counts ranged from 454 to 507, with a median of 493, while RZ counts ranged from 332 to 373, with a median of 349. No X gates were present in the final transpiled circuits, although X was supported by the backend.

The *ibm_miami* circuits were substantially more compact and exhibited even less transpilation variability. Circuit depth was 206 for all 300 executions, while the total gate count ranged from 449 to 450, with a median of 449. Every circuit contained exactly 60 CZ gates and had a two-qubit depth of 60. Each circuit also contained 192 SX operations, while RZ counts ranged only from 197 to 198, with a median of 197. No X, RX, or identity gates were present in the final *ibm_miami* transpiled circuits.

Physical-qubit placement nevertheless varied between executions. Ten distinct final logical-to-physical index layouts were recorded on *ibm_cleveland*, compared with four on *ibm_miami*; the most common *ibm_miami* layout was used in 288 of the 300 executions. Thus, the *ibm_cleveland* repetitions sampled appreciable compiler and layout variability, whereas the *ibm_miami* executions were compiled to an almost invariant circuit structure and predominantly reused the same physical mapping. Overall, both datasets represent repeated executions of the same compact six-qubit logical circuit, while the markedly lower depth and two-qubit-gate burden on *ibm_miami* demonstrate the effect of backend connectivity, native gate support, and layout selection on the hardware resources required to implement the QTF ansatz.

It is worth noting the major difference in QPU usage time. Each *ibm_cleveland* job took approximately 5 seconds, while each *ibm_miami* took 35 seconds, a 7-fold increase in runtime despite the latter having substantially lower transpiled depth and gate count. This difference cannot be attributed to classical transpilation time (1.09 sec for *ibm_cleveland* vs 0.86 sec for *ibm_miami*) and likely reflects backend-specific pulse scheduling, shot-repetition delays, measurement/reset durations, dynamical-decoupling schedules, or fixed execution overhead.

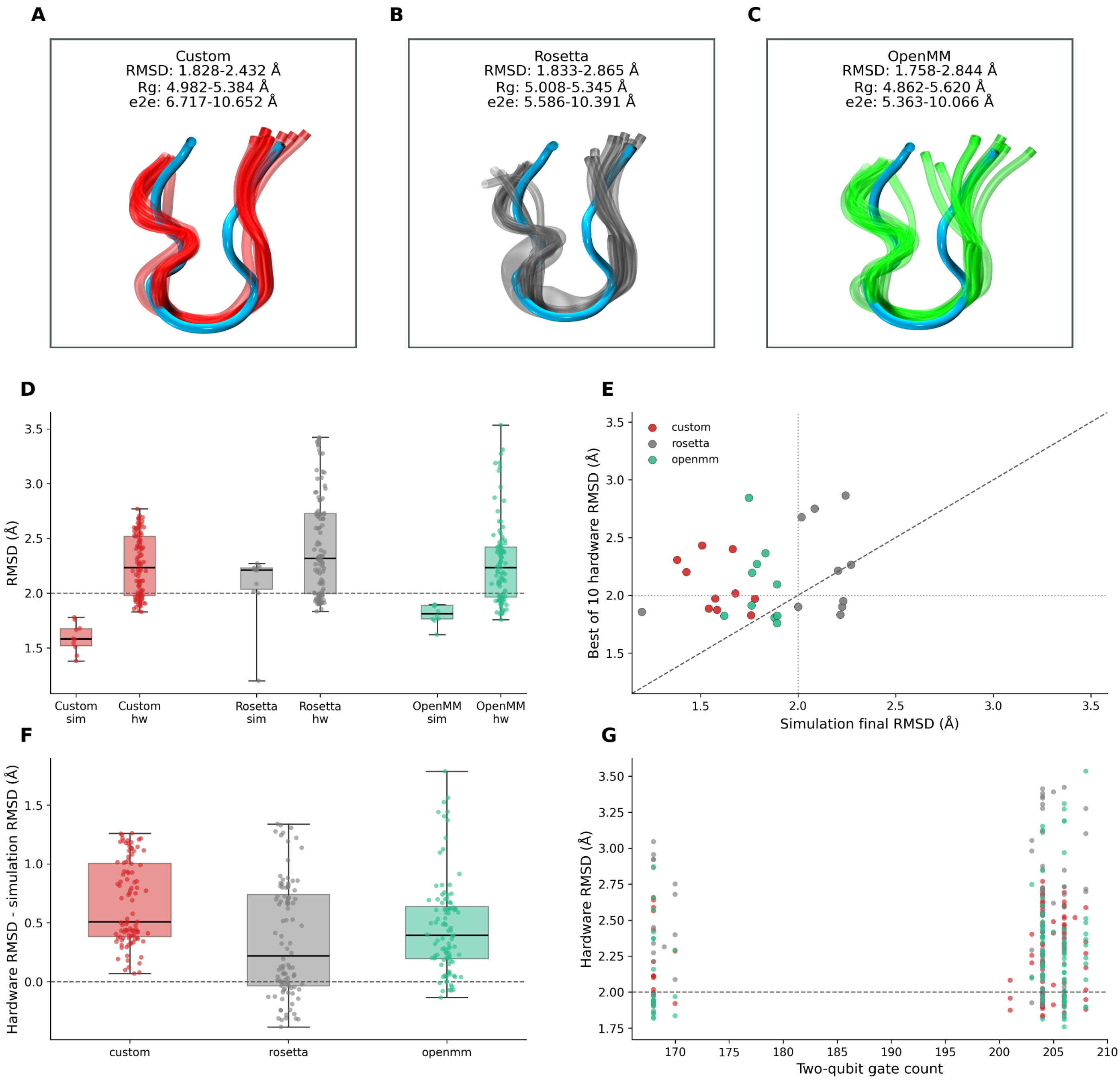

**Figure 6:** Hardware results of 5AWL structure predictions using optimized circuit parameters on *ibm_cleveland*. Panels A-C show structural overlays of the best results from each of the ten selected 5AWL starting hardware replicas for the custom energy function, Rosetta, and OpenMM, respectively. Predicted hardware structures are colored by energy function and overlaid against the experimental structure in cyan; reported ranges correspond to the ten displayed structures. Panel D compares the RMSD distributions of the selected simulation final models with all 100 hardware repeats per energy function. Panel E compares each simulation final model with the best RMSD obtained among its ten matched hardware replicas; points below the diagonal indicate that the hardware CDF-decoded structure has lower RMSD than the corresponding phase-decoded simulation structure, meaning that structural accuracy improved. Panel F shows the per-replica RMSD change relative to the matched simulation model, where negative values indicate improvement and positive values indicate degradation. Panel G compares hardware RMSD with the transpiled two-qubit gate count, showing that gate count alone does not explain the observed structural variability.

The simulation starting structures and hardware structures were produced using different decoders. Simulation structures used gauge-fixed relative statevector phases, whereas hardware structures used the empirical probability CDF mapping. Consequently, differences between the two sets of structures reflect both the change in decoding rule and hardware-related effects such as finite-shot sampling, gate noise, readout noise, mitigation, and transpilation. Because a matched noiseless CDF-decoded baseline was not included, these contributions cannot be separated in the present analysis.

The analysis on *ibm_cleveland* evaluated whether circuit parameters obtained from the best 5AWL simulation models could still recover native-like structures when executed on quantum hardware. Figures 6A-C show structural overlays of the best hardware result obtained from each of the ten selected starting hardware replicas for the custom energy function, Rosetta, and OpenMM, respectively. These overlays therefore represent one structure per “warm started” replica, rather than the ten lowest-RMSD hardware structures pooled across all. This avoids over-representing any single starting model that happened to produce multiple low-RMSD replicas. Across all three energy functions, the overlays retain recognizable 5AWL beta-hairpin-like structure, indicating that the hardware executions preserved meaningful structural information from the encoded circuit parameters. However, the spread of the overlays also shows that hardware execution introduced variability, with some starting replicas producing noticeably displaced conformations even when the best initial simulation model was selected.

Figure 6D compares the RMSD distributions of the original simulation final models with all corresponding warm start hardware outputs. Each energy function includes ten simulation starting models and 100 hardware replicas. The simulation starting models were already highly optimized, especially for the custom energy function, whose selected starts occupied the tightest and lowest-RMSD range. After hardware execution, the RMSD distributions shifted upward for all three energy functions. Of the 300 jobs executed on *ibm_cleveland*, roughly one third recovered native-like 5AWL conformations, defined here as RMSD values below 2.0 Å. Native-like structures were obtained in 29/100 custom energy-function runs, 28/100 Rosetta runs, and 31/100 OpenMM runs. The custom energy function produced hardware RMSDs ranging from 1.828 to 2.769 Å, with a median of 2.233 Å and mean of 2.253 Å. Rosetta produced RMSDs from 1.833 to 3.423 Å, with a median of 2.316 Å and mean of 2.428 Å. OpenMM produced the best individual structure, with a minimum RMSD of 1.758 Å, and had a median of 2.234 Å, mean of 2.273 Å, and maximum of 3.535 Å.

Figure 6E evaluates the best-case hardware outcome for each starting replica. Each point compares the simulation final-model RMSD on the x-axis with the lowest hardware RMSD obtained among its ten hardware replicas on the y-axis. Points below the diagonal indicate cases where hardware execution improved the structure relative to the matched simulation endpoint, whereas points above the diagonal indicate that even the best hardware replica was worse than the original simulation structure. Using this per-starting-replica view, 5/10 starting replicas crossed the 2.0 Å native-like threshold for each energy function. However, improvement relative to the corresponding simulation endpoint was highly cost function dependent: 0/10 custom energy function starts improved, compared with 5/10 Rosetta starts and 3/10 OpenMM starts. This pattern is consistent with the custom energy function starts already being the strongest input set, leaving little room for hardware execution to improve them, while several Rosetta and OpenMM starts began at higher RMSD and therefore had more opportunity to shift toward lower-RMSD conformations.

Figure 6F shows the RMSD change for every hardware repeat relative to its matched simulation starting model, calculated as hardware RMSD minus simulation RMSD. Positive values therefore indicate degradation after hardware execution, while negative values indicate improvement. For the custom energy function, the distribution is almost entirely positive, again showing that hardware execution generally worsened the already accurate starting structures. Rosetta has the broadest distribution and includes the largest number of negative values, consistent with the observation that it had the most improved repeats. OpenMM shows an intermediate behavior, with many repeats worsening relative to the simulation endpoint but a subset improving. Together, Figures 6E and 6F show that repeated hardware execution is useful: although most individual replicas degrade the starting structure, some can recover a lower-RMSD conformation, particularly when the simulation starting model is not already near the best observed accuracy.

Figure 6G examines whether the hardware RMSD is explained simply by the number of two-qubit gates after transpilation. The points are broadly scattered, and similar two-qubit gate counts can produce substantially different RMSDs. This indicates that two-qubit gate count alone is not sufficient to explain the structural accuracy of the hardware outputs. Other factors, including transpilation layout, physical qubit calibration, readout noise, sampling stochasticity, and error mitigation behavior, likely contribute to the observed variability. Overall, the plot shows that the encoded QTF circuit parameters retain native-like structural signals on hardware, but the recovered structure is sensitive to hardware execution effects, making repeated sampling and best-replica selection important for obtaining the most accurate hardware models.

These results indicate that hardware execution on *ibm_cleveland* preserves meaningful structural signal from the saved simulation circuit parameters. The stochastic hardware readout and transpiled circuit layout can shift the reconstructed structure, sometimes improving the selected endpoint and sometimes degrading it. The ten-replica strategy is therefore useful because it samples variability arising from hardware execution, measurement noise, mitigation, and physical qubit layout.

The corresponding *ibm_miami* experiments provided a direct alternate QPU comparison using the same 30 warm-start circuits and ten hardware executions per starting model. Figure S1A-C shows the best *ibm_miami* hardware result obtained from each starting replica, using the same selection applied in Figure 6A-C (the best out of 10 hardware structures per starting model). The resulting overlays retained recognizable beta-hairpin-like structure, but showed greater terminal dispersion than the *ibm_cleveland* structures. Among the ten rendered structures for each energy function, the N-to-C terminal Cα distance ranged from 5.294 to 13.285 Å for the custom energy function, 7.765 to 12.424 Å for Rosetta, and 4.482 to 12.887 Å for OpenMM, compared with 5.509 Å in the experimental 5AWL structure. The corresponding *ibm_cleveland* ranges were 6.717-10.652 Å, 5.586-10.391 Å, and 5.363-10.066 Å, respectively. Thus, the *ibm_miami* ensembles extended to substantially larger terminal separations, particularly for Rosetta and OpenMM. Because this end-to-end metric is defined by the two terminal Cα atoms, whereas the reported RMSD excludes the terminal residues and evaluates the structural core, a model can retain a moderately native-like core while exhibiting pronounced terminal displacement.

The complete *ibm_miami* RMSD distributions were also broader and shifted upward relative to *ibm_cleveland* (Figure S1D). Of the 300 *ibm_miami* jobs, 45 produced structures below the 2.0 Å native-like threshold, compared with 88 of the 300 *ibm_cleveland* jobs. Native-like *ibm_miami* structures were obtained in 21/100 custom energy-function runs, 12/100 Rosetta runs, and 12/100 OpenMM runs. The custom energy function produced RMSDs ranging from 1.819 to 3.157 Å, with a median of 2.301 Å and mean of 2.333 Å. Rosetta produced RMSDs from 1.884 to 3.759 Å, with a median of 2.475 Å and mean of 2.585 Å. OpenMM produced the best individual *ibm_miami* structure, with a minimum RMSD of 1.782 Å, but also showed the widest overall range, reaching 3.950 Å, with a median of 2.293 Å and mean of 2.409 Å. Relative to *ibm_cleveland*, the higher Rosetta and OpenMM maxima indicate larger high-RMSD tails, even though both devices occasionally recovered structures near or below the native-like threshold.

The best-of-ten analysis further showed that repeated execution remained beneficial on *ibm_miami*, although native-like recovery was less frequent than on *ibm_cleveland* (Figure S1E). The best *ibm_miami* repeat crossed the 2.0 Å threshold for 3/10 custom, 2/10 Rosetta, and 4/10 OpenMM starting models, compared with 5/10 for each energy function on *ibm_cleveland*. Improvement relative to the matched simulation endpoint was observed for 0/10 custom, 3/10 Rosetta, and 1/10 OpenMM starting models. The repeat-level RMSD changes in Figure S1F similarly show that most *ibm_miami* executions degraded the simulation endpoint, while a smaller subset, primarily among the Rosetta starts, shifted toward lower RMSD.

Because every *ibm_miami* circuit contained exactly 60 CZ gates, Figure S1G does not provide an internal test of whether two-qubit gate-count variation explains structural accuracy; instead, it shows a wide RMSD distribution at a fixed two-qubit gate burden. Considered together with the *ibm_cleveland* results, the *ibm_miami* experiment confirms that the hardware-compatible decoder can recover native-like conformations on a second QPU architecture, but also demonstrates substantial backend influence. The smaller and nearly invariant *ibm_miami* transpiled circuits did not yield a narrower structural distribution or a higher native-like recovery rate, indicating that transpiled gate count alone is insufficient to predict downstream conformational accuracy. Backend-specific pulse scheduling, calibration, physical-qubit properties, measurement noise, and sampling variability remain important determinants of the recovered structures and support the continued use of repeated execution and best-replica selection.

## DISCUSSION AND CONCLUSIONS

*Ab initio* protein structure prediction remains a challenging task and an important field of research. It has historically been computationally intractable at a larger scale, due to the exponentially growing search spaces. The sampling power of a quantum computer could present new opportunities to further expand these approaches. The vast majority of QPSP efforts employed reduced representations of proteins on discrete lattices, in many ways limiting the biologically realistic modeling of these structures. As quantum hardware capabilities evolve, perhaps it is time to venture into new approaches. Protein structures are largely defined by the angles present along their backbone and sidechains. Our team developed a novel method for encoding protein angles into qubit phases, presenting, to the best of our knowledge, the first ever QPSP approach modeling protein ensembles in continuous space at a full-atom resolution.

**Summary of Findings**

We evaluated QTF on two standard miniature protein benchmarks: chignolin (5AWL; 10 residues, 166 atoms) and Trp-cage (2JOF; 20 residues, 284 atoms), using three scoring backends: a customizable empirical energy function, Rosetta, and OpenMM. Across the full simulation campaign, the intended design was 1,200 replicas per protein, with 400 replicas per energy function. The custom energy function and Rosetta completed all planned replicas for both proteins, while OpenMM produced fewer valid final models because a subset of runs failed during optimization with numerical parameter-binding instabilities. The resulting analysis nevertheless included 2,333 completed replica trajectories with matched snapshot data and more than 7.3 million retained structures across the two proteins.

The clearest structural recovery was observed for chignolin, the smaller beta-hairpin target. Final replica models reached native-like accuracy in a minority of cases, with the best final model reaching 1.199 Å RMSD. However, the retained low-energy snapshot ensembles revealed substantially stronger sampling than the final endpoints alone. Across the replicas, nearly all Rosetta and OpenMM trajectories and all custom energy trajectories contained at least one retained snapshot below the 2.0 Å native-like threshold. The best retained snapshot reached 0.623 Å RMSD, demonstrating that QTF can sample highly accurate near-ground-state conformations even when the final selected model is not the most native-like structure visited during optimization.

Trp-cage was a more difficult benchmark, consistent with its more heterogeneous fold containing helical, turn, and loop elements rather than the simpler chignolin beta-hairpin. No final model or retained snapshot crossed the 2.0 Å native-like threshold under the current energy functions. Even so, snapshot mining improved the best observed RMSD from 3.512 Å among final models to 2.501 Å among retained snapshots, and the per-replica best-snapshot distributions shifted lower than the corresponding final-model distributions for all three energy functions. Thus, even for the harder target, the optimization trajectories frequently visited more native-like conformations than were preserved by the endpoint selection rule.

A central finding is that QTF should be analyzed as a sampling method rather than only as a final-model generator. Each replica retained up to 5,000 low-energy snapshots, with an energy-spacing filter used to avoid filling the ensemble with nearly redundant structures from the same local region of the trajectory. These retained structures are therefore not intended to represent uniform coverage of the full torsional conformational space; they are a selected low-energy subset of the regions explored by each optimization. The value of this design is that it exposes native-like or near-native conformations that would be missed if only the final optimization endpoint were considered.

The folding-funnel and correlation analyses further show that model selection remains the main bottleneck. When sampled structures were rescored on a shared minimized-potential scale (GROMACS using amber99sb-ildn), low-RMSD conformations were often present but were not consistently assigned the lowest energies, and snapshot energy-RMSD Spearman correlations were weak. The raw-energy funnels showed backend-specific landscape structure: the custom energy function concentrated many native-like 5AWL snapshots into a low-RMSD region, whereas Rosetta and OpenMM produced broader low-energy bands that could include both near-native and higher-RMSD conformations. These results indicate that all three scoring approaches can guide sampling toward useful regions of conformational space, but none yet provides a fully reliable ranking function for selecting the most structurally accurate model from the sampled basin.

This energy-RMSD imbalance is not unexpected. Scoring functions used in protein structure prediction are typically evaluated by their ability to enrich native-like conformations relative to decoys across a statistical ensemble, rather than by a strict requirement that the single lowest-energy structure is always the closest to the experimental fold. Therefore, a useful scoring function can still guide sampling toward native-like regions of conformational space while failing to rank the most accurate structure as the absolute minimum. This is a central challenge in force-field development: generalizing across sequences, folds, and conformational states remains difficult, and no universal force field exists that reliably ranks structural accuracy across all contexts. The continued development and parameterization of new energy functions therefore remains essential. The structural-metric analyses support this interpretation. RMSD, radius of gyration, and end-to-end distance together showed that the custom energy function produced final-model ensembles whose median global compactness and chain extension were noticeably closer to the experimental structures for both proteins than the Rosetta and OpenMM final-model ensembles. At the same time, the best retained snapshots from all three energy functions often improved substantially over the final models. Thus, the custom energy function appears to provide the strongest endpoint behavior in these experiments, while snapshot mining reveals that useful conformations are present across all scoring backends.

The *ibm_cleveland* and *ibm_miami* experiments provided a comparative test of whether saved QTF circuit parameters retain useful structural information when transferred to real quantum hardware using the measurement-compatible probability-CDF decoder. For each QPU, we executed the same 30 circuits initialized with parameters from the ten lowest-RMSD 5AWL final models produced by each energy function, with ten hardware executions per starting circuit. On *ibm_cleveland*, 88 of 300 structures remained below the 2.0 Å native-like threshold, compared with 45 of 300 on *ibm_miami*. Best-of-ten selection recovered native-like structures from 5/10 starting models for each energy function on *ibm_cleveland* and from 3/10 custom, 2/10 Rosetta, and 4/10 OpenMM starting models on *ibm_miami*. The best structures reached RMSDs of 1.758 Å on *ibm_cleveland* and 1.782 Å on *ibm_miami*. Although the *ibm_miami* circuits had substantially lower and nearly invariant transpiled depth and two-qubit-gate count, their RMSD distributions were broader, their terminal separations were more variable, and native-like recovery was less frequent. Additionally, the jobs on *ibm_miami* entailed a nearly 7-fold increase in run time, indicating that *ibm_cleveland* was more efficient in allocation usage while also yielding better structural ensembles. Thus, transpiled circuit size alone did not predict structural accuracy; backend-specific calibration, pulse scheduling, physical-qubit properties, measurement noise, and sampling variability also influenced the decoded conformations. Because the simulator and hardware workflows employ mathematically distinct circuit-to-torsion decoders, these experiments represent transfer to a hardware-compatible decoding rule rather than direct recovery of the simulator-derived torsions. Nevertheless, both QPUs recovered native-like structures, and the variability observed across repeated executions supports the use of replicate sampling and best-replica selection in current hardware implementations.

Taken together, these results establish QTF as a continuous-space, all-atom quantum torsional sampling framework with logarithmic qubit scaling in the number of encoded torsional degrees of freedom. The present implementation can already sample near-native chignolin conformations and substantially improve Trp-cage models through trajectory-level snapshot mining. The main remaining challenges are not only qubit count, but also energy-function balance, endpoint selection, physical filtering of strained structures, and the measurement overhead associated with hardware-compatible torsion decoding and any future phase-sensitive recovery protocol. Addressing these issues through improved scoring, post-selection, alternative encodings, and more efficient phase-retrieval protocols will be central to extending the method to larger and more structurally diverse proteins.

**Hardware Feasibility**

While the logarithmic scaling of the QTF architecture presents a profound theoretical advantage in terms of qubit counts, its practical implementation on current physical quantum processors faces significant physical bottlenecks. We must be fully transparent that the core limitation of this approach lies precisely in its internal representation: while continuous torsional angles are phase-encoded within the quantum statevector, the statevector itself is not directly observable on physical Noisy Intermediate-Scale Quantum (NISQ) hardware. On a classical simulator, extracting these phases is a deterministic and trivial mathematical operation. However, because standard computational-basis measurements destroy complex phase information, hardware execution necessitates an alternative mapping strategy derived entirely from observable probability distributions.

Currently, retrieving the necessary angular information on physical devices requires building an empirical probability distribution by executing the quantum circuit for a massive number of shots. Because our hardware-compatible mapping relies on calculating the cumulative distribution function (CDF) from these measured amplitude frequencies, rather than directly reconstructing the encoded phases, the accuracy of the extracted molecular geometry is heavily dependent on the statistical resolution of the measurement vector. In a NISQ environment, hardware noise (including gate fidelity, state preparation and measurement errors, and decoherence) severely distorts these probability distributions. To achieve the precise angular resolution necessary to reconstruct a protein without severe steric clashes, the required number of shots becomes exceptionally high. This leads to substantial computational overhead, heavily limiting the speed and practicality of the method on today’s devices.

Furthermore, as the number of degrees of freedom increases, the state space grows exponentially. This means the shot count must also scale up dramatically to accurately sample the increasingly diluted amplitude probabilities. Under current NISQ limitations, this intensive shot requirement and the resulting noisy readouts effectively trade the problem of qubit quantity for a problem of measurement complexity.

Despite these near-term limitations, there is strong justification for optimism as quantum technology advances. The QTF methodology establishes a robust mathematical and algorithmic foundation built directly for the fault-tolerant era. Future hardware architectures featuring logical, error-corrected qubits will drastically reduce the noise inherent in measurement distributions, lowering the shot overhead required for accurate empirical CDF mapping. Direct recovery of the relative computational-basis amplitude phases used by the simulator decoder would require additional phase-sensitive measurements, such as interference measurements or an appropriate tomographic protocol. These approaches would introduce substantial circuit and measurement overhead, and determining whether they can recover the required relative-phase vector efficiently remains an important direction for future investigation. Alternatively, future hardware implementations could retain a probability-based decoder while investigating more sample-efficient methods for estimating the required probability distribution.

Rather than a preliminary proof-of-concept, the QTF framework delivers a fully realized, scalable architecture. It definitively proves that all-atom, continuous-space protein modeling is not only mathematically viable on a logarithmically scaled register, but practically implementable today. This establishes a powerful standard for hybrid quantum biophysics pipelines that may become more practical on physical QPUs as hardware coherence improves and more efficient probability estimation or phase-sensitive measurement protocols are developed.

**Limitations and Future Directions**

Recognizing both the benefits and limitations presented here, there are new directions worth exploring. First, while the logarithmic scaling of the number of qubits required with respect to the number of angles present in the target protein sequence is highly advantageous, the practical limitations inherent in the employed phase encoding are equally important and present challenges to the scalability, particularly when executing on quantum hardware where the statevector is not directly accessible after measurement. Alternative encoding methods are certainly worth pursuing. Empirical scoring functions also present their own limitations, in particular with generalizability and landscape imbalance. With the rise of quantum chemistry approaches employed on quantum hardware for determining molecular energies, incorporation of such methods for scoring structures derived from the torsional sampling presented here, is an attractive future direction for the field.

## ADDITIONAL INFORMATION

**Availability**

The QTF codebase is released open source under the MIT license and available on GitHub at https://github.com/cumbof/qtf. The PHEAT library, which provides many importable functionalities to QTF, including creating and importing various energy functions, is also open source and available on GitHub at https://github.com/BlankenbergLab/pheat.

Please note that because the Rosetta software suite is not distributed under the MIT license, it cannot be packaged with the main MIT-licensed releases of QTF or PHEAT. To ensure full compliance with Rosetta's licensing agreements, the specific version of QTF that directly integrates the Rosetta energy function (and excludes PHEAT) has been archived in a dedicated branch within the main QTF repository (https://github.com/cumbof/qtf/tree/archive/pre-pheat-main). The code within that specific branch is governed by a license similar to the standard Rosetta license through Rosetta Commons, which is free for academic and non-profit institutions but also available to commercial users for a fee.

The results of the analysis discussed in the paper, including both simulation and hardware execution data for 5AWL and 2JOF, are available at https://doi.org/10.5281/zenodo.22088098.

**Author Contribution**

FC and BR conceived the original exploration of the off-lattice approach. FC implemented the initial algorithmic framework and codebase, and continuously optimized the software. BR developed and optimized the custom energy function and classical post-processing workflow, as well as conducted and analyzed the quantum hardware experiments. BR and VP performed and analyzed the simulations. VP developed the CDF approach for the quantum hardware experiments. NK and JJ provided support in executing preliminary simulations and contributed to interpretations of the findings. DB supervised all aspects of this work and developed the PHEAT code base, which serves as the entry point for all scoring mechanisms in QTF. All authors contributed to the development of the QTF codebase, as well as writing and reviewing this manuscript.

**Conflict of Interests**

No conflicts to disclose.

**Acknowledgments**

We thank Dr. Rui-Hao Li, Research Data Scientist at the Department of Computational Life Sciences, Cleveland Clinic Research, for reviewing our manuscript and providing feedback on the quantum methodology underlying QTF.

We also acknowledge the use of Google's Gemini Pro 3.1 to refine the phrasing, grammar, and overall readability of the manuscript. The use of this language model was strictly limited to linguistic assistance. Claude Code (powered by Anthropic's Opus 4.8 model), and Codex (using OpenAI's GPT-5.5 model) was used for assisting authors at coding tasks. No artificial intelligence was employed in the ideation and methodological design. All scientific concepts, architectures, and conclusions presented herein are entirely the original work of the authors, reflecting their professional expertise.

**REFERENCES**

1. Jumper J, Evans R, Pritzel A, Green T, Figurnov M, Ronneberger O, et al. Highly accurate protein structure prediction with AlphaFold. Nature. 2021;596: 583–589.

2. Abramson J, Adler J, Dunger J, Evans R, Green T, Pritzel A, et al. Accurate structure prediction of biomolecular interactions with AlphaFold 3. Nature. 2024;630: 493–500.

3. Baek M, DiMaio F, Anishchenko I, Dauparas J, Ovchinnikov S, Lee GR, et al. Accurate prediction of protein structures and interactions using a three-track neural network. Science. 2021;373: 871–876.

4. Krishna R, Wang J, Ahern W, Sturmfels P, Venkatesh P, Kalvet I, et al. Generalized biomolecular modeling and design with RoseTTAFold All-Atom. Science. 2024;384: eadl2528.

5. Baek M, Anishchenko I, Humphreys I, Cong Q, Baker D, DiMaio F. Efficient and accurate prediction of protein structure using RoseTTAFold2. bioRxiv. 2023. doi:10.1101/2023.05.24.542179

6. Berman HM, Westbrook J, Feng Z, Gilliland G, Bhat TN, Weissig H, et al. The Protein Data Bank. Nucleic Acids Res. 2000;28: 235–242.

7. Doga H, Raubenolt B, Cumbo F, Joshi J, DiFilippo FP, Qin J, et al. A Perspective on Protein Structure Prediction Using Quantum Computers. J Chem Theory Comput. 2024;20: 3359–3378.

8. Perdomo-Ortiz A, Dickson N, Drew-Brook M, Rose G, Aspuru-Guzik A. Finding low-energy conformations of lattice protein models by quantum annealing. Sci Rep. 2012;2: 571.

9. Babbush R, Perdomo-Ortiz A, O'Gorman B, Macready W, Aspuru-Guzik A. Construction of energy functions for lattice heteropolymer models: A case study in constraint satisfaction programming and adiabatic quantum optimization. arXiv [quant-ph]. 2012. doi:10.48550/arXiv.1211.3422

10. Li R-H, Doga H, Raubenolt B, Mostame S, DiSanto N, Cumbo F, et al. Quantum algorithm for protein structure prediction using the face-centered cubic lattice. arXiv [quant-ph]. 2025. doi:10.48550/arXiv.2507.08955

11. Linn H, Li R-H, Holden A, Saki AA, DiFilippo F, Radivoyevitch T, et al. Efficient quantum protein structure prediction with problem-agnostic ansatzes. arXiv [quant-ph]. 2025. doi:10.48550/arXiv.2509.18263

12. Fingerhuth M, Babej T, Ing C. A quantum alternating operator ansatz with hard and soft constraints for lattice protein folding. arXiv [quant-ph]. 2018. doi:10.48550/arXiv.1810.13411

13. Robert A, Barkoutsos PK, Woerner S, Tavernelli I. Resource-efficient quantum algorithm for protein folding. Npj Quantum Inf. 2021;7: 1–5.

14. Chandarana P, Hegade NN, Montalban I, Solano E, Chen X. Digitized counterdiabatic quantum algorithm for protein folding. Phys Rev Appl. 2023;20. doi:10.1103/physrevapplied.20.014024

15. Wong R, Chang W-L. Fast quantum algorithm for protein structure prediction in hydrophobic-hydrophilic model. J Parallel Distrib Comput. 2022;164: 178–190.

16. Agathangelou A, Manawadu D, Tavernelli I. Quantum algorithm for protein side-chain optimisation: Comparing quantum to classical methods. arXiv [quant-ph]. 2025. doi:10.48550/arXiv.2507.19383

17. Zhang Y, Yang Y, Martin W, Lin K, Wang Z, Lu C-C, et al. A Quantum Framework for Protein Binding-Site Structure Prediction on Utility-Level Quantum Processors. Adv Sci (Weinh). 2026;13: e13641.

18. Zhang Y, Yang Y, Chen F, Lu C-C, Saeidi N, Volchenboum SL, et al. A hybrid quantum-AI framework for protein structure prediction on NISQ devices. arXiv [cs.ET]. 2025. doi:10.48550/arXiv.2510.06413

19. Pamidimukkala JV, Bopardikar S, Dakshinamoorthy A, Kannan A, Dasgupta K, Senapati S. Protein Structure Prediction with High Degrees of Freedom in a Gate-Based Quantum Computer. J Chem Theory Comput. 2024;20: 10223–10234.

20. Geoffrey A S B. Protein structure prediction using AI and quantum computers. bioRxiv. bioRxiv; 2021. doi:10.1101/2021.05.22.445242

21. Shajan A, Kaliakin D, Liang F, Pellegrini T, Doga H, Bhowmik S, et al. Molecular Quantum Computations on a Protein. arXiv [quant-ph]. 2025. doi:10.48550/arXiv.2512.17130

22. Kaliakin D, Shajan A, Liang F, Merz KM Jr. Implicit Solvent Sample-Based Quantum Diagonalization. J Phys Chem B. 2025;129: 5788–5796.

23. Kaliakin D, Shajan A, Liang F, Robledo Moreno J, Li Z, Mitra A, et al. Accurate quantum-centric simulations of intermolecular interactions. Commun Phys. 2025;8: 396.

24. Merz KM, Jr., Shajan A, Kaliakin D, Liang F, Otsuka Y, et al. Crossing the 12,000-atom barrier with heterogeneous quantum-classical supercomputing: quantum chemistry of protein-ligand complexes. arXiv [quant-ph]. 2026. doi:10.48550/arXiv.2605.01138

25. Shajan A, Kaliakin D, Mitra A, Robledo Moreno J, Li Z, Motta M, et al. Toward Quantum-Centric Simulations of Extended Molecules: Sample-Based Quantum Diagonalization Enhanced with Density Matrix Embedding Theory. J Chem Theory Comput. 2025;21: 6801–6810.

26. Li Z, Bazayeva M, Pellegrini T, Bhowmik S, Das S, Kaliakin D, et al. Protein-Ligand Free Energy Perturbation on Quantum Hardware. arXiv [quant-ph]. 2026. doi:10.48550/arXiv.2604.09857

27. Bazayeva M, Li Z, Kaliakin D, Liang F, Shajan A, Das S, et al. Quantum-centric alchemical free energy calculations. arXiv [physics.chem-ph]. 2025. doi:10.48550/arXiv.2506.20825

28. Miyazawa S, Jernigan RL. Residue-residue potentials with a favorable contact pair term and an unfavorable high packing density term, for simulation and threading. J Mol Biol. 1996;256: 623–644.

29. Zeng H, Liu K-S, Zheng W-M. The Miyazawa-Jernigan contact energies revisited. Open Bioinforma J. 2012;6: 1–8.

30. Dill KA. Theory for the folding and stability of globular proteins. Biochemistry. 1985;24: 1501–1509.

31. Lau KF, Dill KA. A lattice statistical mechanics model of the conformational and sequence spaces of proteins. Macromolecules. 1989;22: 3986–3997.

32. Parsons J, Holmes JB, Rojas JM, Tsai J, Strauss CEM. Practical conversion from torsion space to Cartesian space for in silico protein synthesis. J Comput Chem. 2005;26: 1063–1068.

33. AlQuraishi M. Parallelized Natural Extension Reference Frame: Parallelized Conversion from Internal to Cartesian Coordinates. J Comput Chem. 2019;40: 885–892.

34. Kandala A, Mezzacapo A, Temme K, Takita M, Brink M, Chow JM, et al. Hardware-efficient variational quantum eigensolver for small molecules and quantum magnets. Nature. 2017;549: 242–246.

35. Engh RA, Huber R. Accurate bond and angle parameters for X-ray protein structure refinement. Acta Crystallogr A. 1991;47: 392–400.

36. Engh RA, Huber R. Structure quality and target parameters. International Tables for Crystallography. Chester, England: International Union of Crystallography; 2012. pp. 474–484.

37. Alford RF, Leaver-Fay A, Jeliazkov JR, O'Meara MJ, DiMaio FP, Park H, et al. The Rosetta All-Atom Energy Function for Macromolecular Modeling and Design. J Chem Theory Comput. 2017;13: 3031–3048.

38. Chaudhury S, Lyskov S, Gray JJ. PyRosetta: a script-based interface for implementing molecular modeling algorithms using Rosetta. Bioinformatics. 2010;26: 689–691.

39. Eastman P, Swails J, Chodera JD, McGibbon RT, Zhao Y, Beauchamp KA, et al. OpenMM 7: Rapid development of high performance algorithms for molecular dynamics. PLoS Comput Biol. 2017;13: e1005659.

40. Maier JA, Martinez C, Kasavajhala K, Wickstrom L, Hauser KE, Simmerling C. ff14SB: Improving the Accuracy of Protein Side Chain and Backbone Parameters from ff99SB. J Chem Theory Comput. 2015;11: 3696–3713.

41. Zhou H, Zhou Y. Distance-scaled, finite ideal-gas reference state improves structure-derived potentials of mean force for structure selection and stability prediction. Protein Sci. 2002;11: 2714–2726.

42. Miyazawa S, Jernigan RL. Estimation of effective interresidue contact energies from protein crystal structures: quasi-chemical approximation. Macromolecules. 1985;18: 534–552.

43. Powell MJD. A direct search optimization method that models the objective and constraint functions by linear interpolation. Advances in Optimization and Numerical Analysis. Dordrecht: Springer Netherlands; 1994. pp. 51–67.

44. Zhang Z, Ragonneau TM, Schueller J. zaikunzhang/prima: Version 0.5. Zenodo; 2023. doi:10.5281/ZENODO.8052654

45. Abraham MJ, Murtola T, Schulz R, Páll S, Smith JC, Hess B, et al. GROMACS: High performance molecular simulations through multi-level parallelism from laptops to supercomputers. SoftwareX. 2015;1: 19–25.

46. Lindorff-Larsen K, Piana S, Palmo K, Maragakis P, Klepeis JL, Dror RO, et al. Improved side-chain torsion potentials for the Amber ff99SB protein force field. Proteins. 2010;78: 1950–1958.

47. Honda S, Akiba T, Kato YS, Sawada Y, Sekijima M, Ishimura M, et al. Crystal structure of a ten-amino acid protein. J Am Chem Soc. 2008;130: 15327–15331.

48. Akiba T, Ishimura M, Odahara T, Harata K, Honda S. Crystal structure of a mutant of chignolin, cln025. Worldwide Protein Data Bank; 2015. doi:10.2210/pdb5awl/pdb

49. Barua B, Lin JC, Williams VD, Kummler P, Neidigh JW, Andersen NH. The Trp-cage: optimizing the stability of a globular miniprotein. Protein Eng Des Sel. 2008;21: 171–185.

50. Barua B, Andersen NH. The trp-cage: Optimizing the stability of a globular miniprotein. Worldwide Protein Data Bank; 2008. doi:10.2210/pdb2jof/pdb

# Logarithmic-scale variational quantum eigensolver for off-lattice protein structure prediction in continuous torsional angle space

Supplementary Information

Fabio Cumbo, Bryan Raubenolt, Varun Puram, Natalie Katzenmeyer, Jayadev Joshi, Daniel Blankenberg

## S1. Hardware execution on `ibm_miami`

We executed 300 warm-start hardware jobs on `ibm_miami` using the saved final circuit parameters from the ten selected 5AWL starting replicas for each of the custom, Rosetta, and OpenMM energy functions. Each job used the probability-CDF torsion decoder and 8,192 SamplerV2 shots with randomized error suppression. The compiled circuits used six logical QTF qubits, exactly 60 CZ gates, a depth of 206, and 192 SX gates; the total gate count varied only from 449–450 across the 300 executions. The hardware results are summarized below.

The ten displayed custom, Rosetta, and OpenMM structures span RMSD ranges of 1.819–2.573, 1.884–3.227, and 1.782–3.347 Å, respectively. The corresponding e2e ranges are 5.294–13.285, 7.765–12.424, and 4.482–12.887 Å. Thus, the Miami hardware ensemble shows a wider structural spread than the Cleveland ensemble, especially for end-to-end distance and the Rosetta/OpenMM RMSDs, despite the substantially more compact transpiled circuits. This comparison reinforces that backend execution time and structural variability are not determined by circuit depth or two-qubit count alone.

## S2. Residue-specific side-chain torsional degrees of freedom

QTF assigns side-chain torsional variables according to residue templates and hard caps rather than by introducing every chemically definable terminal dihedral. The limits avoid optimization variables that are redundant because of molecular symmetry or constrained by rigid planar geometry. In particular, the terminal guanidinium group of arginine does not require an additional independent terminal torsion, while aromatic side chains retain the torsions needed to position the ring but do not assign independent rotations within the rigid planar ring system. Glycine and alanine have no side-chain $\chi$ variables. The map below shows the default `chi_mode=all` variables after these residue-specific reductions; backbone $\phi$, $\psi$, and (under the default windowed-$\omega$ treatment) peptide-link $\omega$ variables are separate from the side-chain counts.

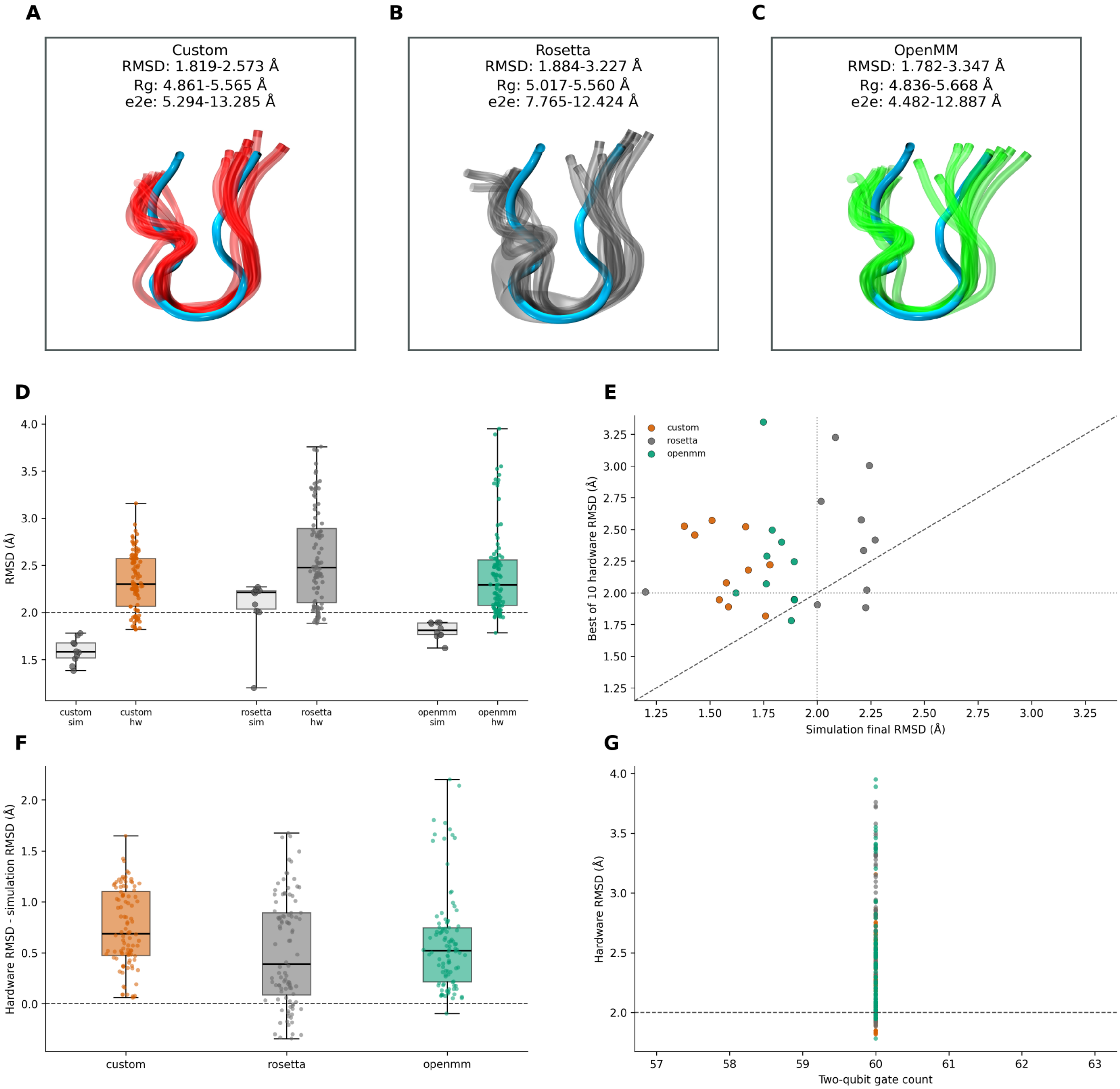

Figure S1: **Structural and quantitative outcomes of hardware execution on `ibm_miami`.** (A–C) Structural overlays of the best hardware repeat from each of the ten selected 5AWL starting replicas for the custom, Rosetta, and OpenMM energy functions, respectively. The experimental structure is shown in cyan and the hardware-derived models are shown as transparent overlays. Reported RMSD, radius of gyration ($R_g$), and end-to-end distance (e2e) ranges are calculated only from the ten rendered structures in each panel. (D) RMSD distributions for the ten simulation-derived starting models ("sim") and all 100 corresponding hardware repeats ("hw") for each energy function. (E) RMSD of the best of ten hardware repeats for each starting replica plotted against its simulation final-model RMSD; the diagonal denotes equal RMSD, and dotted lines mark the 2.0 Å native-like threshold. (F) Change in RMSD relative to the corresponding simulation starting model for all hardware repeats, with positive values indicating structural degradation and negative values indicating improvement. (G) Hardware-model RMSD as a function of transpiled two-qubit gate count across all 300 executions. The horizontal dashed lines in D and G indicate the 2.0 Å native-like threshold.

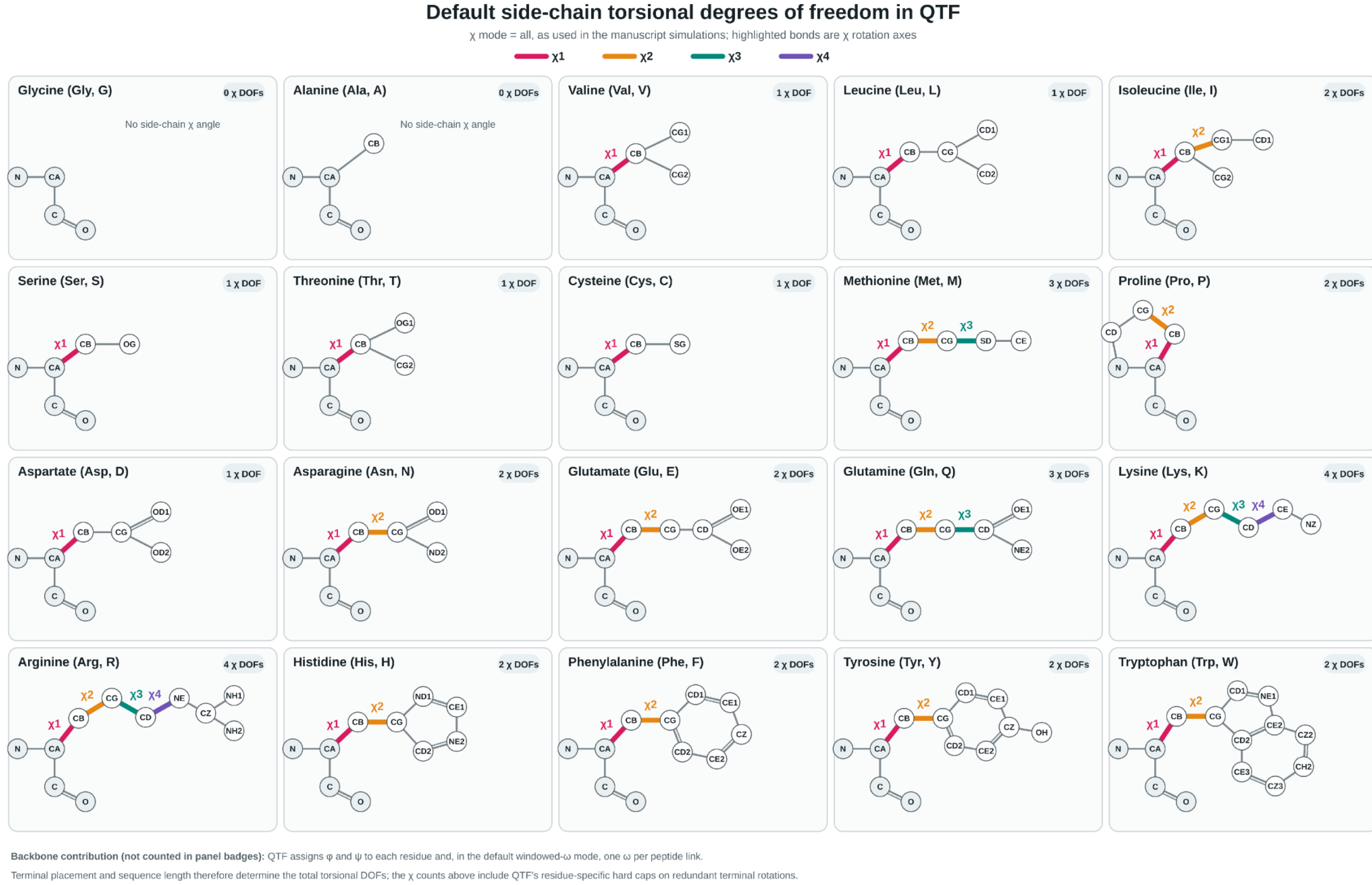

Figure S2: **Residue-specific side-chain torsional degrees of freedom used in QTF.** Heavy-atom schematics are shown for the 20 standard amino acids, with the rotation axes exposed under `chi_mode=all` highlighted and labeled $\chi_1$–$\chi_4$. The badge in each panel gives the number of side-chain $\chi$ degrees of freedom used by QTF for that residue. The displayed counts include the residue templates and residue-specific hard caps and therefore do not necessarily include every chemically definable terminal dihedral. Symmetry and planar geometry are used to avoid wasting degrees of freedom; for example, the terminal guanidinium group of arginine is symmetry-equivalent at its end, and aromatic rings are rigid and planar.

## S3. Custom effective-energy derivation

The complete derivation and implementation-level definitions of the QTF custom effective energy are included verbatim from the accompanying source file. This objective is a classical conformational scoring function evaluated after circuit readout; it is not a qubit-Hamiltonian expectation value.

# Supplementary Methods: QTF Custom Effective Energy

QTF evaluates each circuit-generated conformation by reconstructing Cartesian coordinates from internal torsional degrees of freedom and scoring the resulting structure with a classical effective Hamiltonian, defined here as an empirical potential-energy objective over protein conformational coordinates. This objective is evaluated classically after circuit readout; it is not a qubit Hamiltonian expectation value.

$$E_{\text{total}} = E_{\text{constraint}} + E_{\text{elec}} + E_{\text{hbond}} + E_{\text{burial}} + E_{\text{rama}} + E_{\text{rotamer}} + E_{\text{geom}} + E_{\omega} + E_{\pi\text{-stack}} + E_{\text{steric}} + E_{\text{disulfide}}. \quad (1)$$

The steric and peptide-bond terms are grouped from implementation-level subterms:

$$E_{\text{steric}} = E_{\text{vdW}} + E_{\text{clash}}, \quad (2)$$

$$E_{\text{vdW}} = E_{\text{vdW,rep}} + E_{\text{vdW,att}}, \quad (3)$$

$$E_{\text{clash}} = E_{\text{hard clash}} + E_{\text{adjacent heavy}}, \quad (4)$$

$$E_{\omega} = E_{\omega,\text{center}} + E_{\omega,\text{window}}. \quad (5)$$

## Definitions

Let $r_{ij}$ be the distance between atoms $i$ and $j$, $q_i$ the assigned effective partial charge, $R_i$ the assigned van der Waals radius, and $\epsilon_i$ the atom-type Lennard-Jones well-depth parameter. Let $\mathcal{N}$ denote the masked nonbonded atom-pair set, $\mathcal{N}_{1-4}$ the reduced-weight 1–4 pair set, and $\Delta(a,b) = ((a - b + \pi) \bmod 2\pi) - \pi$ the wrapped angular difference.

| Symbol | Meaning |
|---|---|
| $N_{\text{res}}$ | Number of residues in the modeled sequence |
| $r_{ij}$ | Cartesian distance between atoms $i$ and $j$ in Å |
| $q_i$ | Effective partial charge assigned to atom $i$ |
| $R_i$ | van der Waals radius assigned to atom $i$ |
| $\epsilon_i$ | Atom-type Lennard-Jones well-depth parameter assigned to atom $i$ |
| $\gamma_i$ | Hydrophobicity value assigned from the residue containing atom $i$ |
| $\mathbf{1}[\cdot]$ | Indicator function equal to 1 when the condition is true and 0 otherwise |
| $\text{clip}(x, a, b)$ | Value $x$ bounded to the interval $[a, b]$ |
| $\mathcal{N}$ | Masked longer-range nonbonded atom-pair set |
| $\mathcal{N}_{1-4}$ | Masked 1–4 nonbonded atom-pair set evaluated with reduced weight |
| $\mathcal{H}$ | Hydrophobic atom set used by the burial proxy |
| $\Delta(a, b)$ | Wrapped angular difference between torsions $a$ and $b$ |

### End-to-End Constraint

For terminal C$\alpha$ atoms separated by $d_{\text{ee}}$, QTF uses a mild sequence-length-aware restraint during stages in which the end-to-end constraint is active:

$$d_{\text{target}} = 4.5 + 0.40 \max(0, N_{\text{res}} - 5), \quad (6)$$

$$s_{\text{ee}} = 1.5 + 0.05 N_{\text{res}}, \quad (7)$$

$$\delta_{\text{ee}} = \max\left(0, |d_{\text{ee}} - d_{\text{target}}| - s_{\text{ee}}\right), \quad (8)$$

$$E_{\text{constraint}} = w_{\text{ee}}\, k_{\text{stage}}\, \delta_{\text{ee}}^2. \quad (9)$$

The default $w_{\text{ee}}$ is 1.0. The stage-dependent coefficient is $k_{\text{stage}} = 8.0$ in early high-force stages and 1.5 in the natural-relaxation stage. Here $d_{\text{target}}$ is the sequence-length-dependent target end-to-end distance, $s_{\text{ee}}$ is a slack tolerance around that target, and $\delta_{\text{ee}}$ is the distance violation after the slack region is removed.

## Hydrophobic Burial

The implementation reports this term as `sasa`, but it is a hydrophobic-burial proxy rather than a literal solvent-accessible surface-area calculation. For hydrophobic atoms $i \in \mathcal{H}$,

$$n_i = \sum_j \frac{1}{1 + \exp(r_{ij} - 6.0)} - 1, \tag{10}$$

$$b_i = \operatorname{clip}\left(\frac{n_i}{35.0}, 0, 1\right), \tag{11}$$

$$A_i^{\text{exposed}} = 30.0(1 - b_i), \tag{12}$$

$$E_{\text{burial}} = w_{\text{burial}} \sum_{i \in \mathcal{H}} \gamma_i A_i^{\text{exposed}}, \tag{13}$$

where $\gamma_i$ is the residue hydrophobicity assigned to atom $i$. The default $w_{\text{burial}}$ is 0.7. Here $n_i$ is a soft neighbor count around hydrophobic atom $i$, $b_i$ is the resulting burial fraction, and $A_i^{\text{exposed}}$ is an effective exposed hydrophobic area. Larger positive $\gamma_i$ values make exposed hydrophobic atoms more costly.

## Hydrogen Bonding

For each backbone nitrogen donor, a virtual amide hydrogen $H$ is placed from local peptide geometry. Carbonyl oxygen acceptors from residues separated by at least two sequence positions are evaluated:

$$E_{\text{hbond}} = w_{\text{hbond}} \sum_{H,O} -50 \exp\left[-\frac{(r_{HO} - 2.0)^2}{0.5}\right] 2\left(|\cos\theta_{NHO}| - 0.4\right) \mathbf{1}[r_{HO} < 3.5]\mathbf{1}[\cos\theta_{NHO} < -0.4]. \tag{14}$$

Here $\theta_{NHO}$ is the N–H–O angle as implemented from the normalized H→N and H→O vectors. The default $w_{\text{hbond}}$ is 0.75. The term is negative for accepted donor–acceptor pairs, so favorable hydrogen-bond-like geometries lower the total score. The radial Gaussian is centered at an H–O distance of 2.0 Å, while the two indicator functions enforce the 3.5 Å distance cutoff and directional criterion.

## Electrostatics

Electrostatics use a Coulomb-like term over nonbonded charged pairs, with a 1.0 Å distance floor and a uniform dielectric of 4.0:

$$E_{\text{elec}} = \sum_{(i,j) \in \mathcal{N},\, |q_i q_j| > 10^{-4}} \frac{332.0637\, q_i q_j}{4.0 \max(r_{ij}, 1.0)}. \tag{15}$$

The factor 332.0637 converts charges in elementary charge units and distances in Å to kcal mol$^{-1}$. The denominator 4.0 is the uniform implicit dielectric used by the current custom scorer.

## Disulfide Term

For cysteine sulfur atoms $S_i$, plausible disulfide-like contacts are rewarded near 2.05 Å and sulfur over-saturation is penalized:

$$E_{\text{disulfide,reward}} = -25 \sum_{i<j} \exp\left[-\frac{(r_{S_i S_j} - 2.05)^2}{0.5}\right] \mathbf{1}[r_{S_i S_j} < 3.0], \tag{16}$$

$$\sigma_i = \sum_{j \neq i} \exp\left[-\frac{(r_{S_i S_j} - 2.05)^2}{0.5}\right] \mathbf{1}[r_{S_i S_j} < 3.0], \tag{17}$$

$$E_{\text{disulfide,overload}} = 40 \sum_i (\sigma_i - 1)^2\, \mathbf{1}[\sigma_i - 1 > 0.1], \tag{18}$$

$$E_{\text{disulfide}} = E_{\text{disulfide,reward}} + E_{\text{disulfide,overload}}. \tag{19}$$

This term is zero unless at least two cysteine SG atoms are present. Here $r_{S_i S_j}$ is the distance between cysteine sulfur atoms $S_i$ and $S_j$, and $\sigma_i$ is the effective number of close disulfide-like partners for sulfur $S_i$. The overload penalty discourages one sulfur from being rewarded for multiple simultaneous disulfide-like contacts.

## van der Waals Packing

For nonbonded and reduced 1–4 heavy-atom pairs, QTF defines an effective contact distance

$$
\begin{aligned}
r_{ij}^{\min} &= 0.95(R_i + R_j), && (20)\\
\sigma_{ij}^{\mathrm{LJ}} &= \frac{r_{ij}^{\min}}{2^{1/6}}, && (21)\\
\epsilon_{ij} &= \sqrt{\epsilon_i \epsilon_j}, && (22)\\
r'_{ij} &= \max(r_{ij}, 1.2), && (23)
\end{aligned}
$$

and evaluates

$$L_{ij} = 4\epsilon_{ij} \left[ \left( \frac{\sigma_{ij}^{\mathrm{LJ}}}{r'_{ij}} \right)^{12} - \left( \frac{\sigma_{ij}^{\mathrm{LJ}}}{r'_{ij}} \right)^{6} \right]. \tag{24}$$

The repulsive branch is softened:

$$L_{ij}^{\mathrm{rep}} = \begin{cases} \max(L_{ij}, 0), & \max(L_{ij}, 0) \leq 25, \\ 25 + \log\left(1 + \max(L_{ij}, 0) - 25\right), & \max(L_{ij}, 0) > 25, \end{cases} \tag{25}$$

and the attractive branch is clipped:

$$L_{ij}^{\mathrm{att}} = \mathrm{clip}(L_{ij}, -2.5, 0). \tag{26}$$

Thus,

$$
\begin{aligned}
E_{\mathrm{vdW,rep}} &= 0.01 \sum_{(i,j)\in\mathcal{N}} L_{ij}^{\mathrm{rep}} + 0.01(0.35) \sum_{(i,j)\in\mathcal{N}_{1-4}} L_{ij}^{\mathrm{rep}}, && (27)\\
E_{\mathrm{vdW,att}} &= 0.10 \sum_{(i,j)\in\mathcal{N}} L_{ij}^{\mathrm{att}} + 0.10(0.35) \sum_{(i,j)\in\mathcal{N}_{1-4}} L_{ij}^{\mathrm{att}}. && (28)
\end{aligned}
$$

Here $r_{ij}^{\min}$ is the effective preferred contact distance for an atom pair and is set to 0.95 times the sum of the assigned van der Waals radii. The symbol $\sigma_{ij}^{\mathrm{LJ}}$ is the Lennard-Jones sigma parameter derived from that preferred contact distance using $r_{\min} = 2^{1/6}\sigma$. The pair well depth $\epsilon_{ij}$ is the geometric mean of atom-type well depths. The floored distance $r'_{ij}$ prevents singular behavior for severe overlaps. The 0.35 factor reduces the contribution of 1–4 pairs, while 0.01 and 0.10 are the default repulsive and attractive scale factors.

## Hard and Adjacent-Residue Clash Penalties

The hard-clash wall penalizes very short nonbonded heavy-atom distances:

$$E_{\mathrm{hard\ clash}} = 5000 \sum_{(i,j)\in\mathcal{N}} \left[ \frac{\max(0, 1.20 - r_{ij})}{1.20} \right]^4 + 5000(0.25) \sum_{(i,j)\in\mathcal{N}_{1-4}} \left[ \frac{\max(0, 1.20 - r_{ij})}{1.20} \right]^4. \tag{29}$$

For adjacent residues $r$ and $r+1$, all heavy-atom pairs are checked except the peptide C–N bond. With

$$T_{ij} = \max\left(1.35, 0.55(R_i + R_j)\right), \tag{30}$$

the local adjacent-residue penalty is

$$E_{\mathrm{adjacent\ heavy}} = 10 \sum_{i\in r,\, j\in r+1} \left[ \frac{\max(0, T_{ij} - r_{ij})}{0.50} \right]^2. \tag{31}$$

Here $T_{ij}$ is the minimum allowed local heavy-atom distance for adjacent residues. It is the larger of 1.35 Å and 55% of the atom-pair van der Waals radius sum. The 0.50 Å denominator sets the width of the local quadratic wall.

## Rotamer Prior

For side-chain torsions, QTF applies residue-specific chi-angle priors. For beta-branched residues V, I, and T:

$$E_{\chi_1}^{\mathrm{V/I/T}} = -3\left[\exp\left(-\frac{\Delta(\chi_1,\pi)^2}{0.5}\right) + \exp\left(-\frac{\Delta(\chi_1,-\pi/3)^2}{0.5}\right)\right]. \tag{32}$$

For proline:

$$E_{\chi_1}^{\mathrm{P}} = 10\min\left[\Delta(\chi_1,-0.5)^2, \Delta(\chi_1,0.5)^2\right]. \tag{33}$$

For aromatic residues W, F, Y, and H:

$$E_{\chi_1}^{\mathrm{aro}} = -2\left[\exp\left(-\frac{\Delta(\chi_1,\pi)^2}{0.45}\right) + 0.8\exp\left(-\frac{\Delta(\chi_1,-\pi/3)^2}{0.45}\right) + 0.8\exp\left(-\frac{\Delta(\chi_1,\pi/3)^2}{0.45}\right)\right]. \tag{34}$$

For other chi1 values:

$$E_{\chi_1}^{\mathrm{other}} = 1 + \cos(3\chi_1). \tag{35}$$

For later chi angles, with centers $\mathcal{C} = \{-\pi/3, \pi/3, \pi\}$:

$$E_{\chi_2}^{\mathrm{aro}} = -1.5\sum_{c\in\mathcal{C}}\exp\left[-\frac{\Delta(\chi_2,c)^2}{0.35}\right], \tag{36}$$

$$E_{\chi 2+}^{\mathrm{other}} = -0.75\sum_{c\in\mathcal{C}}\exp\left[-\frac{\Delta(\chi,c)^2}{0.50}\right]. \tag{37}$$

The implemented rotamer contribution is the sum over available chi torsions, multiplied by the default scale $w_{\mathrm{rotamer}} = 1.0$. All torsions are in radians. The centers $-\pi/3$, $\pi/3$, and $\pi$ correspond to gauche−, gauche+, and trans-like rotamer wells. The wrapped difference $\Delta$ avoids discontinuities at the $-\pi/\pi$ boundary.

## Aromatic $\pi$ Stacking

For aromatic residues F, Y, and W, ring centroids $c_i$ and approximate normals $n_i$ are computed from available ring atoms. For aromatic pairs within 7.0 Å,

$$E_{\pi\text{-stack}} = \sum_{i<j}\left[-4\exp\left(-(d_{ij}-5.0)^2\right)\mathbf{1}[|n_i\cdot n_j|<0.3]\mathbf{1}[4.5<d_{ij}<6.0] - 5\exp\left(-(d_{ij}-3.8)^2\right)\mathbf{1}[|n_i\cdot n_j|>0.8]\mathbf{1}[3.4<d_{ij}<4.5]\right] \tag{38}$$

where $d_{ij} = ||c_i - c_j||$. The default scale is 1.0. Here $c_i$ is the centroid of the selected ring atoms for aromatic residue $i$, $n_i$ is an approximate ring normal, $d_{ij}$ is the centroid–centroid distance, and $|n_i \cdot n_j|$ distinguishes approximately T-shaped from parallel aromatic geometries.

## Ramachandran Prior

For each residue with defined $\phi$ and $\psi$, QTF uses Gaussian wells near alpha-helical and beta-sheet regions:

$$d_\alpha = (\phi+1.0)^2 + (\psi+0.8)^2, \tag{39}$$

$$d_\beta = (\phi+2.3)^2 + (\psi-2.4)^2. \tag{40}$$

For glycine, left-handed wells are also allowed:

$$d_{\alpha L} = (\phi-1.0)^2 + (\psi-0.8)^2, \tag{41}$$

$$d_{\beta L} = (\phi-2.3)^2 + (\psi+2.4)^2, \tag{42}$$

$$E_{\mathrm{rama}}^{\mathrm{Gly}} = -3\exp\left[-\frac{\min(d_\alpha, d_\beta, d_{\alpha L}, d_{\beta L})}{0.6}\right]. \tag{43}$$

For non-glycine residues,

$$d_{\mathrm{forbidden}} = (\phi+2.0)^2 + (\psi-1.0)^2, \tag{44}$$

$$E_{\mathrm{rama}}^{\mathrm{nonGly}} = -3\exp\left(-\frac{d_\alpha}{0.6}\right) - 3\exp\left(-\frac{d_\beta}{0.6}\right) + 5\exp\left(-\frac{d_{\mathrm{forbidden}}}{1.0}\right). \tag{45}$$

The $\phi$ and $\psi$ torsions are in radians. Negative terms reward proximity to idealized alpha-helical or beta-sheet regions, while the positive term penalizes a representative disfavored region for non-glycine residues. Glycine receives symmetric left-handed wells because of its greater backbone flexibility.

### Peptide-Bond Omega Restraint

In current main-branch runs, the default omega mode is `window`. Let

$$\omega_{\min} = 170^\circ, \quad \omega_{\max} = 190^\circ, \quad \omega_0 = 180^\circ, \quad h_\omega = \frac{\omega_{\max} - \omega_{\min}}{2}. \tag{46}$$

In window mode, a raw circuit angle $x \in [-\pi, \pi]$ for an omega DOF is mapped to

$$\omega_{\text{raw}} = \omega_{\min} + \text{clip}\left(\frac{x+\pi}{2\pi}, 0, 1\right)(\omega_{\max} - \omega_{\min}). \tag{47}$$

The effective omega used for reconstruction is then bounded to the trans window:

$$\omega_{\text{eff}} = \text{clip}\left(\omega_{\text{raw}}, \omega_{\min}, \omega_{\max}\right), \tag{48}$$

with signed trans values near $-180^\circ$ first converted to the equivalent positive representation.
The center restraint is

$$E_{\omega,\text{center}} = w_\omega \sum_{k=1}^{N_{\text{res}}-1} \left(\frac{\omega_{\text{eff},k} - \omega_0}{h_\omega}\right)^2, \tag{49}$$

with default $w_\omega = 1.0$. The window-violation component is

$$E_{\omega,\text{window}} = w_{\omega,\text{window}} \sum_k v(\omega_{\text{raw},k})^2, \tag{50}$$

with default $w_{\omega,\text{window}} = 25.0$, where $v(\omega)$ is the normalized distance outside the accepted signed trans windows $[170^\circ, 190^\circ]$ and $[-180^\circ, -170^\circ]$. Here $x$ is the raw circuit-derived torsional value before omega-specific mapping, $\omega_{\text{raw}}$ is the corresponding proposed peptide-bond torsion, and $\omega_{\text{eff}}$ is the bounded torsion used for Cartesian reconstruction. In `fixed` mode omega is held at trans and is not optimized; in `free` mode the raw torsion is not mapped into the trans window before the window penalty is evaluated.

### Geometry Integrity

The geometry term includes chirality and peptide-planarity checks. For a residue with CA, N, C, and CB atoms,

$$V = [(N - CA) \times (C - CA)] \cdot (CB - CA), \tag{51}$$

$$E_{\text{chirality}} = 50(1 - V)^2 \mathbf{1}[V < 1]. \tag{52}$$

For peptide planes defined by $\text{CA}_r$–$\text{C}_r$–$\text{N}_{r+1}$–$\text{CA}_{r+1}$, let $n_1$ and $n_2$ be the normalized adjacent plane normals. The planarity penalty is

$$E_{\text{planarity}} = 20\,(1 - |n_1 \cdot n_2|)\,\mathbf{1}\,[1 - |n_1 \cdot n_2| > 0.05]. \tag{53}$$

Thus,

$$E_{\text{geom}} = \sum_r E_{\text{chirality},r} + \sum_r E_{\text{planarity},r}. \tag{54}$$

Proline ring closure is handled by the rebuild path and is neutral in the current scoring term. Here $V$ is the signed scalar triple product around the C$\alpha$ stereocenter. The planarity term compares normals from adjacent peptide-plane fragments and penalizes non-planar twists only when the deviation exceeds the 0.05 threshold.

## Implementation Notes

Several coefficients can be overridden by environment variables in the code. The equations above list the default values used by the current implementation. Energies from this custom objective are effective arbitrary scoring units unless otherwise noted; the electrostatic prefactor uses kcal mol$^{-1}$ Å e$^{-2}$ internally, while the full objective combines heterogeneous empirical terms for optimization and ranking.